\documentclass[%
aps,
prx,
reprint,
superscriptaddress,
nofootinbib,
amsmath,
amssymb
]{revtex4-2}
\usepackage{
    physics,
    graphicx,
    multirow,
    tabularx,
    mathtools, 
    placeins, 
    nicefrac, 
    hyperref, 
    threeparttable, 
    siunitx,
    enumitem,
    makecell,
    amsthm,
    amscd
}

\newtheorem{lemma}{Lemma}
\usepackage[capitalize]{cleveref}
\Crefname{section}{Sec.}{Secs.}

\usepackage[dvipsnames]{xcolor}
\hypersetup{
    colorlinks,
    linkcolor={blue!50!black},
    citecolor={blue!50!black},
    urlcolor={blue!80!black}
}

\usepackage[caption=false]{subfig} 
\usepackage{algpseudocode}
\usepackage{algorithm}

\makeatletter
\newcommand\fs@nobottomruled{\def\@fs@cfont{\bfseries}\let\@fs@capt\floatc@ruled
  \def\@fs@pre{\hrule height.8pt depth0pt \kern2pt}%
  \def\@fs@post{}
  \def\@fs@mid{\kern2pt\hrule\kern2pt}%
  \let\@fs@iftopcapt\iftrue}
\makeatother

\floatstyle{nobottomruled}
\restylefloat{algorithm}

\begin{document}


\title{Low-overhead Magic State Circuits with Transversal CNOTs}

\newcommand{\qmaddress}{\affiliation{Quantum Motion, 9 Sterling Way, London N7 9HJ, United Kingdom}}
\newcommand{\oxddress}{\affiliation{Department of Engineering Science, University of Oxford, Parks Road, Oxford OX1 3PJ, United Kingdom}}

\author{Nicholas Fazio}
\thanks{These authors contributed equally to this work. Corresponding authors \url{nicholas.fazio@sydney.edu.au}, \url{mark.webster@ucl.ac.uk}, \url{cai.zhenyu.physics@gmail.com}.}
\affiliation{Centre for Engineered Quantum Systems, School of Physics,\\ The University of Sydney, Sydney, New South Wales 2006, Australia.}

\author{Mark Webster}
\thanks{These authors contributed equally to this work. Corresponding authors \url{nicholas.fazio@sydney.edu.au}, \url{mark.webster@ucl.ac.uk}, \url{cai.zhenyu.physics@gmail.com}.}
\affiliation{Department of Physics \& Astronomy, University College London, London, WC1E 6BT, United Kingdom}

\author{Zhenyu Cai}
\oxddress
\qmaddress

\date{\today}

\begin{abstract}
With the successful demonstration of transversal CNOTs in many recent experiments, it is the right moment to examine its implications on one of the most critical parts of fault-tolerant computation -- magic state preparation. Using an algorithm that can recompile and simplify a circuit of consecutive multi-qubit phase rotations, we manage to construct fault-tolerant circuits for $\ket{CCZ}$, $\ket{CS}$ and $\ket{T}$ states with much lower CNOT depths and qubit counts than before and minimal $T$-depth for the given workspace. These circuits can play crucial roles in fault-tolerant computation with transversal CNOTs, and we hope that the algorithms and methods developed in this paper can be used to further simplify other protocols in similar contexts.
\end{abstract}

\maketitle

\section{\label{sec:intro}Introduction}

For quantum computers to realise their full potential, they must achieve fault tolerance at large scales, and the associated overheads need to be kept low to ensure reasonable device sizes and compute time.
There have been several recent experimental demonstrations of quantum error correction (QEC) across a variety of architectures~\cite{postlerDemonstrationFaulttolerantUniversal2022, bluvsteinLogicalQuantumProcessor2024, ryan-andersonHighfidelityTeleportationLogical2024,googleQECBelowThreshold2024}, substantiating that QEC does work in practice.
The capabilities and limitations of these architectures are quite different, meaning that their respective approaches to large-scale fault tolerance will differ as they utilise schemes that take advantage of the strengths of their architecture. 

One critical component of fault tolerance is the generation of high-quality non-Clifford resources, the so-called magic states, which have been the subject of concerted research.
The resources required for preparing and distilling magic states have been massively reduced through a series of schemes~\cite{litinskiMagicStateDistillation2019,chamberlandVeryLowOverhead2020, itogawa2024ZeroLevelDistillation,gidney2024magicstatecultivationgrowing}, with recent work suggesting that the generation of $\ket{T}$ states can have lower overheads than lattice-surgery logical \text{CNOT}s~\cite{gidney2024magicstatecultivationgrowing} for a certain range of target error rates.

In many cases the non-Clifford resources generated, namely $\ket{T}$ states, must be further synthesised into other non-Clifford gates like CCZ to be implemented in a circuit. Addressing this need has been met with a series of approaches in both gate synthesis and state distillation.
An early example is Jones' fault-tolerant construction for implementing \text{CCZ} from eight $T$ gates using techniques in synthesis~\cite{jonesLowoverheadConstructionsFaulttolerant2013, matroid}.
Later the connections between the phase polynomials of synthesis~\cite{selingerQuantumCircuitsDepth2013} and the triorthogonal matrices of magic state distillation~\cite{Bravyi2012MSDTriorthogonal} were identified, and unified under the portmanteau of synthillation~\cite{campbell2017unified}, which combines the state distillation and gate synthesis steps to further suppress error rates.
Jones' fault-tolerant construction was then recognised as an example of a synthillation scheme with code parameters [\![8,3,2]\!]. Re-expressing extant gadgets has also been used to derive magic measurements for making schemes more accessible~\cite{litinskiMagicStateDistillation2019}, or catalysed state factories for reducing overheads~\cite{gidneyEfficientMagicState2019}. 

When considering explicit implementation and optimisation of schemes for magic state distillation and the closely related synthillation, most of the prior works focus on 2D qubit connectivity using lattice surgery.
On the other hand, significant experimental progress has been made in multiple hardware platforms such as trapped ions and neutral atoms~\cite{postlerDemonstrationFaulttolerantUniversal2022, ryan-andersonHighfidelityTeleportationLogical2024,bluvsteinLogicalQuantumProcessor2024,rodriguezExperimentalDemonstrationLogical2024} in demonstrating additional qubit connectivity and transversal CNOT gates available between code patches.
These capabilities have significant potential to reduce the spacetime overhead of fault-tolerant computation~\cite{zhouAlgorithmicFaultTolerance2024} and naturally, significant room for improvements to magic state preparation circuits in this context. 

In this work, considering the availability of transversal CNOTs and the enhanced connectivity between logical qubits, we have devised an algorithm for optimising \text{CNOT}+$T$ magic state preparation circuits, achieving low qubit counts, low CNOT depth and maximal $T$ parallelisation. The algorithm leverages SWAP gates to reduce the CNOT counts, absorbing SWAPs into the state initialisation round.
For architectures where arbitrary SWAP gates are cheap, this compilation strategy can be extended to more general settings.
The algorithm adopts a greedy approach which can be adapted for different requirements, for example, optimising for \text{CNOT} count rather than \text{CNOT} depth. 
Our work is complementary to recent results in magic state preparation~\cite{gidney2024magicstatecultivationgrowing} since all our circuits start with states prepared in $\ket{T}$, which inherently lends itself to better schemes for high-quality magic states.
Moreover, the required output error rates for schemes preparing $\ket{T}$ do not need to be as low because synthillation further improves the error rates for the final non-Clifford gate.

We start this paper by introducing the fault-tolerant \text{CCZ} scheme of Jones~\cite{jonesLowoverheadConstructionsFaulttolerant2013} in \cref{sec:ccz_distill} as a motivating example, describing its components and the ways that this scheme can be reformulated. In \cref{sec:parallelisation_algo} we introduce the concept of phase rotation operators and outline our algorithm for simplifying blocks of \text{CNOT}s. At the end of the section and also in \cref{sec:MSD-circuits}, we apply our algorithm to construct low overhead fault-tolerant schemes for \text{CCZ}, \text{CS} and $T$ as shown in \cref{fig:CCZCircFinal,fig:CS-rearranged,fig:OurTgate}. This is followed by a brief resource analysis of our \text{CCZ} circuit in \cref{sec:analaysis} and further discussion in \cref{sec:discussion}.

\section{Motivating example: fault-tolerant \text{CCZ}} \label{sec:ccz_distill}
\begin{figure}[htbp]
    \centering
    \includegraphics[width = 0.35\textwidth]{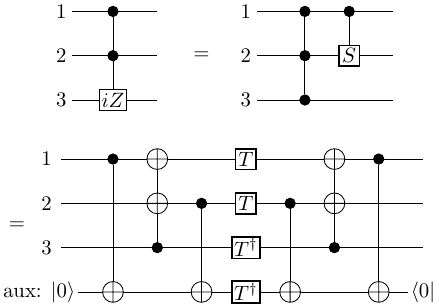}
    \caption{Circuit for implementing \text{CC($i$Z)}.}
    \label{fig:CCiZfromT}
\end{figure}

\begin{figure}[htbp]
    \centering
    \includegraphics[width = 0.35\textwidth]{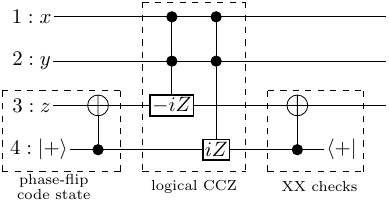}
    \caption{Circuit for fault-tolerant \text{CCZ}. The \text{CCZ}s here can be implemented using \cref{fig:CCiZfromT} and its conjugate, thus there are two additional auxiliary qubits used that are not shown here.}
    \label{fig:CCZRepCode}
\end{figure}

Selinger~\cite{selingerQuantumCircuitsDepth2013} proposed a way to implement \text{CC($i$Z)} (which is equivalent to $\text{CS} \cdot\text{CCZ}$) using a circuit of only $T$-depth one as shown in \cref{fig:CCiZfromT}. Single $Z$ errors on any $T$ and $T^\dagger$ gates in this circuit will propagate to qubit $3$. Hence, Jones~\cite{jonesLowoverheadConstructionsFaulttolerant2013} proposed encoding qubit $3$ into a phase-flip repetition code depicted in \cref{fig:CCZRepCode}, which can detect single $Z$ errors on any $T$ and $T^\dagger$, suppressing the failure rate of the \text{CCZ} from $p$ to $p^2$. This circuit can then be used for preparing the \text{CCZ} resource state $\ket{CCZ}$ by inputting $\ket{+}$ on the first three qubits. For the rest of the paper, we will often use $T$ gates to refer to both $T$ and $T^\dagger$, which should be obvious from the context.

\begin{figure}[htbp]
    \centering
    \includegraphics[width = 0.47\textwidth]{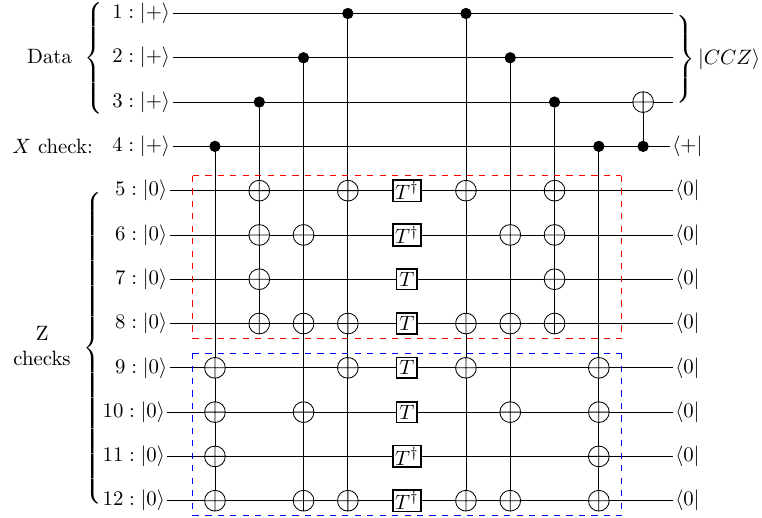}
    \caption{Circuit for distilling \text{CCZ} states from $T$/$T^\dagger$ gates. The red and blue dashed boxes denote the $T$ gates that are used for implementing a particular \text{CC($i$Z)} gate.}
    \label{fig:JonesCirc}
\end{figure}

\begin{figure}[htbp]
    \centering
    \includegraphics[width = 0.47\textwidth]{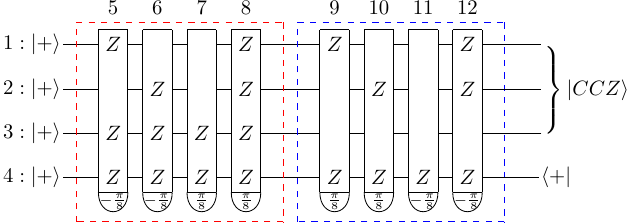}
    \caption{Circuit for distilling \text{CCZ} states using multi-qubit $\pi/8$~phase rotations. The index over each phase rotation denotes which $T$~gate qubit it corresponds to in \cref{fig:JonesCirc}.}
    \label{fig:CCZPhaseRot}
\end{figure}

The explicit circuit constructed by Jones for $\ket{CCZ}$ preparation is shown in \cref{fig:JonesCirc}, with additional auxiliary qubits performing $Z$ checks on the $T$~gates, detecting $X$ errors. These auxiliary qubits can also be used to facilitate lattice surgery~\cite{gidneyEfficientMagicState2019}. Let us focus on only the $T$~gate error suppression power of the circuit, ignoring the memory error and Clifford gate errors and considering only the error of the $T$~gates. In this case, no $X$~error will occur for $T$ gates implemented using gate teleportation, and thus the $Z$~checks will act trivially and will not contribute to the error suppression.
Hence, to reduce the qubit overhead of the circuit while keeping the same $T$~gate error suppression power, these $Z$-check auxiliary qubits can be removed by permuting all the $T$~gates through the \text{CNOT}s, resulting in a series of multi-qubit phase rotations as shown in \cref{fig:CCZPhaseRot}~\cite{litinskiGameSurfaceCodes2019}. Here multi-qubit phase rotations (or $Z$-parity rotations) are simply unitaries generated by tensor products of~$Z$, which is rigorously defined later in \cref{sec:parallelisation_algo}.
In this form, the raw action of $T$~gates on the input states can be understood without worrying about the specific placement of $T$~gates or Clifford gates.
However, all of the $T$~gates have now become multi-qubit $\pi/8$ rotations that cannot be parallelised, leading to maximum $T$-depth. In the next section, we will discuss how to reduce the $T$-depth without introducing auxiliary qubits, maintaining the small qubit count of the circuit.

\section{Algorithms for Phase Rotation Parallelisation}\label{sec:parallelisation_algo}
Similar to the fault-tolerant \text{CCZ} circuit in the last section, more general magic state distillation circuits can also be mapped into a series of phase rotation operators. In this section, we will detail the algorithm for converting a series of phase rotation operators to circuits composed of a qubit permutation followed by alternating blocks of the following two types:
\begin{itemize}
    \item A series of \text{CNOT} gates; and
    \item Application of (possibly different) powers of $T$ in parallel on each of the qubits.
\end{itemize}
The result of the algorithm is a circuit which has minimum $T$-depth without introducing auxiliary qubits, as well as low 2-qubit gate count and depth for the \text{CNOT} circuit blocks. 

Our method is similar to the one set out in \cite{matroid} to reduce $T$-depth, but introduces a new algorithm for synthesising \text{CNOT} Clifford operators. This method expresses the Clifford operator as a circuit that consists of a qubit permutation followed by a series of two-qubit \text{CNOT} gates. After \text{CNOT} synthesis, we commute the permutation for each block through to the beginning of the circuit. The qubit permutation and initial \text{CNOT} block can typically be absorbed into the state preparation part of the circuit. 

The structure of this section is as follows. We first introduce a notation for phase rotation operators and CNOT circuit blocks. We then show how to parallelise a set of independent phase rotation operators by conjugating a series of $T$ operators by a \text{CNOT} Clifford operator. We then introduce our \text{CNOT} synthesis algorithm and finish with an example that illustrates the construction of a circuit for CCZ synthillation.

\subsection{Notation for Phase Rotation and CNOT Operators}

The \textbf{phase rotation operator} is defined as:
\begin{align}
    Z^{\theta}_{\mathbf{u}}  := \exp(i\theta(I - Z(\mathbf{u}))) \propto \exp[ - i \theta Z(\mathbf{u})].
\end{align}
where $\theta \in \mathbb{R}$ is an angle of rotation, $\mathbf{u}$ is a binary vector of length $n$ and $Z(\mathbf{u}):= \prod_{0 \le i < n}Z_i^{\mathbf{u}[i]}$ is the tensor product of $Z$ operators that acts non-trivially on the qubits where $\mathbf{u}[i] = 1$.
The phase rotation operator $Z^{\theta}_{\mathbf{u}}$ has the following action on basis elements $\ket{\mathbf{e}}$:
\begin{align}
Z^{\theta}_{\mathbf{u}}\ket{\mathbf{e}} &= e^{i2\theta (\mathbf{u}\cdot\mathbf{e}\mod 2)}\ket{\mathbf{e}}.\label{eq:phase_rot_action}
\end{align}
We consider a series of phase rotation operators $\prod_{0 \le i < m} Z^{\theta_i}_{\mathbf{u}_i}$ to be \textbf{independent} if the associated binary matrix $U$ with columns $\mathbf{u}_i$ is full-rank, i.e. the set of vectors $\{\mathbf{u}_i\}_{0\leq i < n}$ are linearly independent.
We will also use $Z_i^{\theta}$ to denote a single-qubit rotation that acts non-trivially on qubit $i$. 

We can express any CNOT circuit block as a Clifford operator $\textit{CX}(U)$ where $U$ is an invertible binary matrix referred to as a \textbf{parity matrix} (see \cite{CNOT_parity_matrix}). The operator $\textit{CX}(U)$ 
has symplectic matrix form $\left[\begin{array}{c|c}U&0\\\hline0&(U^T)^{-1}\end{array}\right]$ and 
has the following action on the computational basis element $\ket{\mathbf{e}}$:
\begin{align}
\textit{CX}(U)\ket{\mathbf{e}} 
&=  \ket{U\mathbf{e}}.\label{eq:CX_action}
\end{align}

\subsection{Parallelisation of Phase Rotation Operators}
A set of independent phase rotations is equivalent to a series of $T$ operators conjugated by a CNOT Clifford operator as set out in the following result which is implied by the results in \cite{matroid}:

\begin{lemma}[parallelisation of independent phase rotation operators]\label{thm:phase_rot_par}
Let $\prod_{0 \le i < n}Z^{\theta_i}_{\mathbf{u}_i}$ be a series of independent phase rotation operators on $n$ qubits and let $U$ be the invertible binary matrix with columns $\mathbf{u}_i$. The product of the phase rotation operators can be written as:
\begin{align*}
\prod_{0 \le i < n} Z^{\theta_i}_{\mathbf{u}_i} = \textit{CX}(U)^{-1} \left(\prod_{0 \le i < n} Z^{\theta_i}_{i}\right) \textit{CX}(U).
\end{align*}
\end{lemma}
\textbf{Proof:}
For the computational basis element $\ket{\mathbf{e}}$, $\textit{CX}(U)\ket{\mathbf{e}} = \ket{U\mathbf{e}}$ (see \Cref{eq:CX_action}).
Bit $i$ of $U\mathbf{e}$ is given by $\mathbf{u}_i \cdot \mathbf{e}$ and so $Z^{\theta_i}_i$ applies a phase of $e^{i2\theta_i} \iff \mathbf{u}_i \cdot \mathbf{e} = 1 \mod 2$. This is the same phase applied by the phase rotation operator $Z^{\theta_i}_{\mathbf{u}_i}$ to $\ket{\mathbf{e}}$ (see \Cref{eq:phase_rot_action}). 

Applying the same logic to all $0 \le i < n$, we have:
\begin{align*}
    \prod_{0 \le i < n} Z^{\theta_i}_i \ket{U\mathbf{e}} = \prod_{0 \le i < n}e^{i2\theta_i (\mathbf{u}\cdot\mathbf{e}\mod 2)}\ket{U\mathbf{e}}.
\end{align*}
Noting that $\textit{CX}(U)^{-1}\ket{U\mathbf{e}}=\ket{\mathbf{e}}$, the result follows.

\subsection{Partition of Phase Rotations into Independent Sets}\label{sec:partition_rot}
We have shown how to express a product of $n$ independent phase rotation operators as a \text{CNOT} Clifford operator conjugating a parallel application of $T$ operators on the qubits. In general, we have more than $n$ phase rotation operators to simplify. In this case, we separate the phase rotations into a series of independent sets of rotations which correspond to a series of invertible binary matrices $U_i$, then apply the above result to each~$U_i$. In~\cite{matroid}, a polynomial-time algorithm is presented for finding independent sets of phase rotations using partitions of matroids. For our optimisation, we sample from all possible orderings of the phase rotations. We separate each ordering into sets of size $n$. We reject those which do not form independent sets, then simplify the \text{CNOT} circuit blocks as much as possible to find an optimal solution.

\subsection{Simplification of \text{CNOT} Circuit Blocks}

In this section, we describe an algorithm for simplifying a \text{CNOT} Clifford operator of the form $\textit{CX}(U)$ for some invertible binary matrix $U$. There may be many ways of expressing the operator as a circuit. Our objective is to express the operator in terms of an initial qubit permutation, followed by as few 2-qubit \text{CNOT} gates as possible. 

There is a significant amount of prior work on synthesis of CNOT circuits, aimed at minimisation of CNOT count. 
One approach is to perform Gaussian elimination by column operations on the parity matrix representation of the CNOT circuit \cite{CNOT_parity_matrix}.
As the parity matrix is invertible, the result will be the identity.
Adding column $i$ to column $j$ corresponds to applying a $\text{CNOT}_{ij}$ gate but the resulting circuit may be far from optimal in terms of CNOT count.
In \cite{Patel_Markov_Hayes_2008} the authors present a heuristic algorithm based on splitting the parity matrix into blocks, eliminating common elements within the blocks, then performing Gaussian elimination. 
The authors show that their algorithm is asymptotically optimal, but for small circuits the number of CNOT gates may be far from optimal.
In \cite{schaeffer2014costminimizationapproachsynthesis} various scalar heuristics are used to synthesise CNOT circuits. The authors note that their algorithm may become trapped in local minima and revert to Gaussian elimination when this occurs. 

The greedy algorithm for CNOT synthesis presented here has two main differences to previous approaches. 
Firstly, our method assumes that qubit permutations can be applied freely at the start of the circuit and so we reduce the parity matrix to a permutation matrix rather than the identity. 
This is appropriate for magic state distillation circuits where initial qubit permutations can be absorbed into state preparation. 
Secondly, we use a vector heuristic rather than a scalar heuristic and this reduces the impact of local minima. 
The greedy algorithm results in circuits with fewer CNOT gates than the method in \cite{Patel_Markov_Hayes_2008} because the compilation is up to qubit permutations. 
Benchmarking in a follow-up work~\cite{webster2025heuristicoptimalsynthesiscnot} demonstrates that, for CNOT circuits on 6 or fewer qubits, the greedy algorithm is fast and close to optimal in terms of CNOT count.
The examples in this paper are available at \url{https://github.com/m-webster/CliffordOpt/blob/main/examples/CNOT\%2BT.py}.

Our greedy CNOT synthesis algorithm looks at all possible $\text{CNOT}_{ij}$ operations at each step, selecting the one which minimises the vector of column and row-sums of the resulting matrix. 
In a permutation matrix, the sum of every row and every column is one so the algorithm terminates when an all-ones vector is reached.
A detailed description is set out in \Cref{alg:CNOT_Synthesis}.

\begin{algorithm}[H]
\caption{CNOT Synthesis}\label{alg:CNOT_Synthesis}
\end{algorithm}
\vspace{-4ex}
\begin{algorithmic}
\State\textbf{Input:} 
\State An invertible $n\times n$ binary matrix $U$.
\State\textbf{Output:} 
\State A permutation matrix $P$ plus a series of row operations opList which generate $U$ when applied to $P$.
\State\textbf{Method:} 
\Comment{P will be modified by the algorithm and be returned as the permutation matrix} 
\State Set P := transpose(U) 
\Comment{list of $\text{CNOT}_{ij}$ as ordered pairs (i,j)} 
\State Set opList := [] 
\Comment{if P is a permutation matrix then sum(P) == n}
\State done := (sum(P) == n)
\While{not done}
\Comment{col and row-sums of matrix after applying $\text{CNOT}_{ij}$}
\State wij = []

\For{i in range(n)}
\For{j in range(n)}

\If{i != j}
\State B := copy(P)
\State B[j] := B[i] + B[j]
\State w := sorted(colSums(B) + rowSums(B)) 
\State wij.append((w,i,j))
\EndIf
\EndFor
\EndFor
\Comment{choose (i,j) with min col and row sums after $\text{CNOT}_{ij}$}
\State{w,i,j = min(wij)}
\Comment{update P and opList}
\State P[j] := P[i] + P[j] 
\State opList.append((i,j)) 
\Comment{Check if P is a permutation matrix}
\State done := (sum(P) == n)
\EndWhile
\State return P, reversed(opList)
\end{algorithmic}

\subsection{Fault-tolerant CCZ Example}

Let us now look back to the question of turning the series of phase rotations in the \text{CCZ} synthillation circuit of
\cref{fig:CCZPhaseRot} into a circuit with maximally parallelised $T$ without auxiliary qubits. There are $8$ phase rotations acting on $4$ qubits, hence, in order to parallelise them we need to first divide them into linearly independent sets as discussed in \cref{sec:partition_rot}. By sampling all possible permutations, we find the following reordering of the phase rotation operators of \cref{fig:CCZPhaseRot}:
\\
\\
\includegraphics[width = 0.43\textwidth]{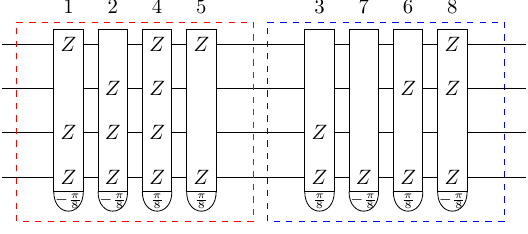}
\\
\\
We partition the phase rotations into two blocks corresponding to the following invertible binary matrices:
\begin{align*}
U_0&:=\begin{bmatrix}
1&0&1&1\\
0&1&1&0\\
1&1&1&0\\
1&1&1&1
\end{bmatrix};
&U_1&:=\begin{bmatrix}
0&0&0&1\\
0&0&1&1\\
1&0&0&0\\
1&1&1&1
\end{bmatrix}.
\end{align*}
Using \Cref{thm:phase_rot_par}, we can write the phase rotations as \text{CNOT} Clifford operators conjugating parallel applications of $T$ and $T^\dag$ operators:
\\
\\
\includegraphics[width = 0.47\textwidth]{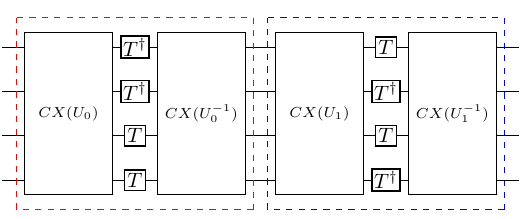}
\\
\\
To simplify the \text{CNOT} blocks, we first group together adjacent blocks as follows:
\\
\\
\includegraphics[width = 0.47\textwidth]{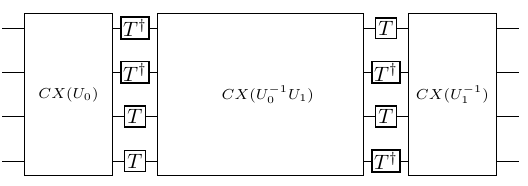}
\\
\\
Applying \cref{alg:CNOT_Synthesis}, we synthesize each CNOT block into a qubit permutation followed by a series of \text{CNOT} gates: 
\\
\\
\includegraphics[width = 0.47\textwidth]{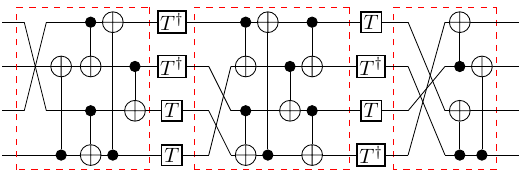}
\\
\\
We commute all qubit permutations through to the beginning of the circuit by updating the \text{CNOT} gates and reordering the $T/T^\dag$ operators. In this case, the permutations cancel but this is not always the case.
We obtain the following:
\\
\\
\includegraphics[width = 0.47\textwidth]{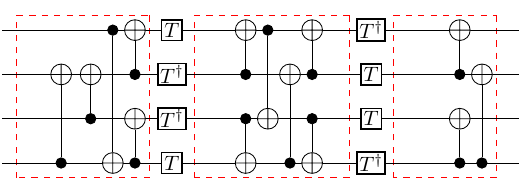}
\\
\\
Absorbing the initial CNOT block and any residual qubit permutations into state preparation using the methods of \Cref{sec:circ_compilation_tricks}, and adding the final gates we derive the following circuit for fault-tolerant CCZ:
\\
\\
\includegraphics[width = 0.47\textwidth]{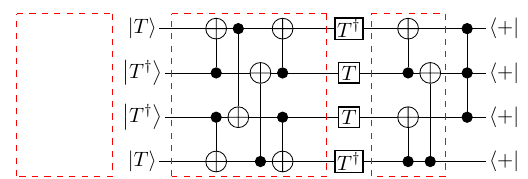}
\\
\\

\begin{figure}[htbp]
    \centering
    \includegraphics[width = 0.47\textwidth]{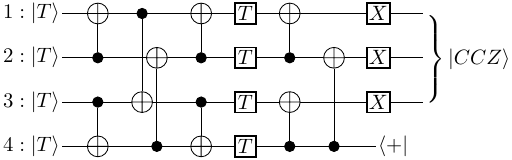}
    \caption{Circuit for fault-tolerant \text{CCZ}. Note that the first layer of \text{CNOT}s only has depth $3$. The CNOT gates in the circuit can be implemented using 2D nearest-neighbour connectivity.}
    \label{fig:CCZCircFinal}
\end{figure}

Using $X\ket{T} \sim \ket{T^\dagger}$ and $X T^\dagger X \sim T$, we can further rewrite the circuit above purely in terms of $T$, CNOTs and $X$ as shown in \cref{fig:CCZCircFinal}.

\section{\label{sec:MSD-circuits}More magic state circuits}

\subsection{\label{subsec:CS}Fault-tolerant CS gate}

Let us now apply our CNOT synthesis algorithm from \cref{sec:parallelisation_algo} to a new example with a similar structure: a fault-tolerant CS gate.

\begin{figure}[t]
    \centering
    \includegraphics[width = 0.3\textwidth]{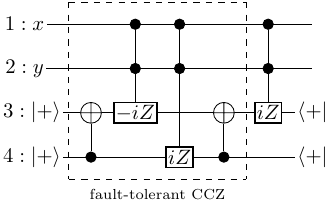}
    \caption{Circuit for fault-tolerantly implementing a \text{CS} with faulty $T$~gates. The CC($i$Z)s here are implemented using \cref{fig:CCiZfromT} and its conjugate.}
    \label{fig:CS-FT}
\end{figure}

In \cref{sec:ccz_distill} we discussed the decomposition $\text{CC($i$Z)} = \text{CS} \cdot\text{CCZ}$ and how this was used by Jones~\cite{jonesLowoverheadConstructionsFaulttolerant2013} to construct a \text{CCZ} gate where any single $T$ gate fault is detectable. In a similar way we derive a circuit that implements a \text{CS} gate that can tolerate single $T$ gate faults shown in \cref{fig:CS-FT}.
This circuit may be understood as the fault-tolerant \text{CCZ} circuit of \cref{fig:CCZRepCode} with an additional \text{CC($i$Z)} applied.
Since the action on the third qubit is the identity it can be used to detect errors from the extra \text{CC($i$Z)}, making the overall scheme fault-tolerant.

Following \cref{sec:ccz_distill}, we may express the circuit of \cref{fig:CS-FT} in terms of 12 parallel $T$/$T^\dagger$, and then further convert that circuit into phase rotations. After regrouping the resultant phase rotations into independent sets we arrive at the following circuit\footnote{This sequence of 12 phase rotations corresponds to the [\![12,2,2]\!] code of Webster et al.~\cite{Webster2023diagonallogical}, where transversal $T$ implements a logical CS gate, or the $NCS$ synthillation protocol of Campbell and Howard for $N=1$~\cite{campbell2017unified}.}
\\
\\
\includegraphics[width = 0.47\textwidth]{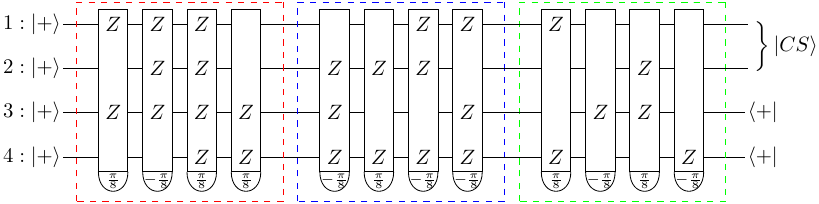}.
\\
\\

\begin{figure}[htbp]
    \centering
    \includegraphics[width = 0.47\textwidth]{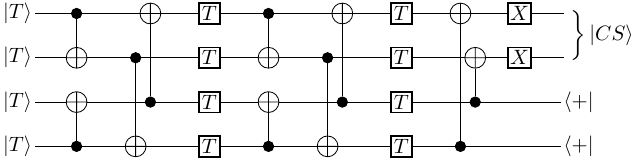}
    \caption{Circuit for distilling CS states rearranged to use in-place parallel $T/T^\dag$ gates.}
    \label{fig:CS-rearranged}
\end{figure}

We then apply the algorithm of \cref{sec:parallelisation_algo}, and the sequence of phase rotations is transformed into alternating layers of \text{CNOT} circuits and transversal $T$/$T^\dagger$. SWAP gates and the initial \text{CNOT}s are then absorbed into the initial $\ket{+}$ states so that the first $T$/$T^\dagger$ layer can be incorporated as part of state initialisation. We arrive at the fault-tolerant \text{CS} circuit of \cref{fig:CS-rearranged} where each CNOT layer has depth 1 or 2. 

\subsection{T gate distillation circuit}\label{subsec:T}

\begin{figure}[htbp]
    \centering
    \includegraphics[width = 0.47\textwidth]{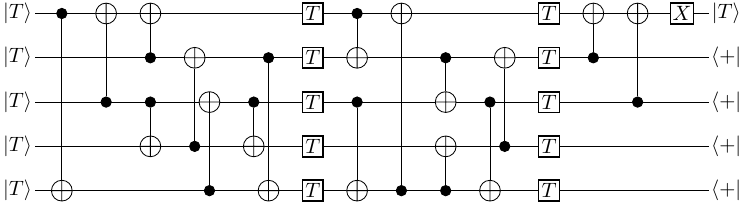}
    \caption{Our simplified circuit for distilling T states. Its CNOT depth is $5+4+2 = 11$
    .}
    \label{fig:OurTgate}
\end{figure}

Another application example is the well-studied $15$-to-$1$ $T$ state distillation scheme derived from the [\![15,1,3]\!] Reed-Muller code~\cite{bravyiUniversalQuantumComputation2005}, which can be implemented using 24 logical patches with lattice surgery as depicted in Ref.~\cite{fowlerLowOverheadQuantum2019}.
As in previous sections, we can turn the circuit into a series of phase rotation operators as outlined by Litinski~\cite{litinskiGameSurfaceCodes2019}, and these rotations can be grouped into linearly independent sets as shown below:
\\
\\
\includegraphics[width = 0.47\textwidth]{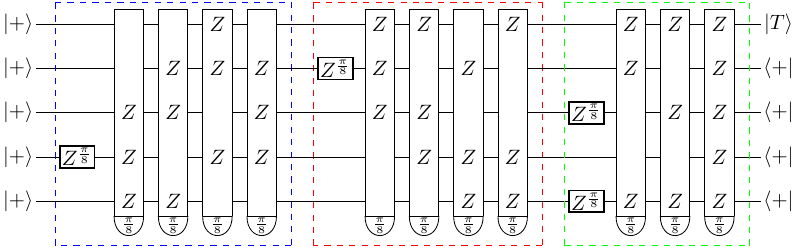}
\\
\\
Once again following the steps outlined in \cref{sec:parallelisation_algo}, each group here can be implemented using parallel $T$~gates and towers of \text{CNOT}s. With further simplifications, we can obtain \cref{fig:OurTgate}.

\begin{figure}[t]
    \centering
    \includegraphics[width =0.47 \textwidth]{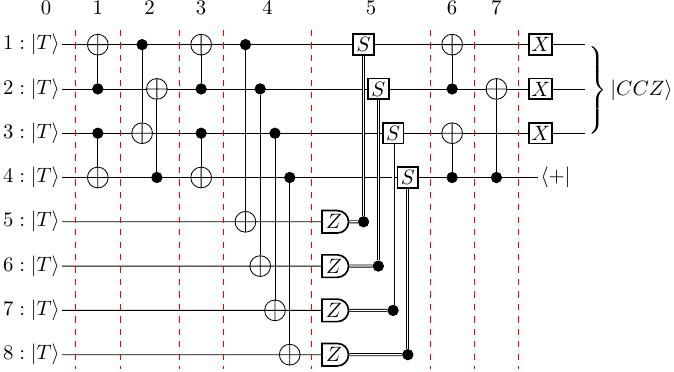}
    \caption{Implementation of a fault-tolerant CCZ with explicit $T$-gadgets. Each round can be implemented using transversal CNOTs, with the exception of round 5. The amount of time $t_\text{s}$ necessary for round 5 is dependent on decoding time, among other factors.}
    \label{fig:impl-circ}
\end{figure}

\begin{figure}[b]
    \centering
    \includegraphics[width =0.47 \textwidth]{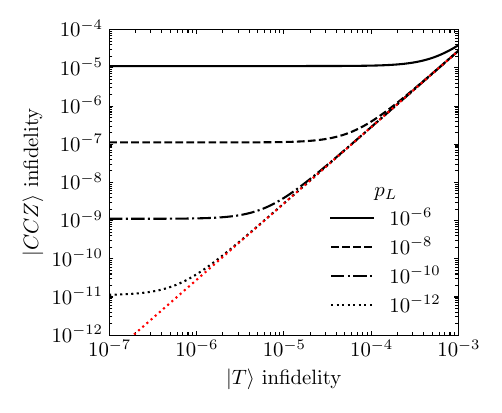}
    \caption{State infidelity of an output CCZ state using~\cref{fig:impl-circ}, taking $t_\text{s}=1$. In round 0, $\ket{T}$ states are implemented with $Z$ errors at rate $p_T\gg p_L$. Thereafter, at the end of each round each logical qubit experiences depolarising noise, including auxiliary qubits, at rate $p_L$. In red is $p_\text{CCZ}= 28p_T^2$, marking the ideal Clifford case of $p_L=0$.}
    \label{fig:err-rates}
\end{figure}

\section{\label{sec:analaysis}Resource Analysis}

In this section we compare the spacetime overhead of our $\ket{CCZ}$ synthillation circuit to the $\ket{CCZ}$ factories of~\cite{gidneyEfficientMagicState2019}, using surface codes as the base code.

As shown in \cref{fig:impl-circ}, we need $8$ workspace qubits and we will use $4$ more qubits for the $S$ state preparation and parallel teleportation of the $S$ correction, thus we use $12$ logical qubits in total. 
On the other hand, by taking the most optimistic case for the scheme in Ref.~\cite{gidneyEfficientMagicState2019}, it still needs at least $18$ logical qubits for implementation because of the auxiliary patches used in lattice surgery.
$T$~states must be prepared in-place when using only 18 logical qubits, which means that different instances of the circuit cannot be interleaved efficiently, and so the time overhead will be $8.5d$.
For our circuit in \cref{fig:impl-circ}, we see that $7$ rounds of transversal gates are needed (an additional round of CNOTs from $S$~gate teleportation and Pauli gates are implemented using Pauli frame updates, shown in \cref{fig:CCZIncludeS}).
We will attach one round of stabiliser checks after each round of transversal gates to deal with the noise introduced
(one round should be sufficient since the noise introduced by transversal gates is less than a single round of checks, which consists of 4 rounds of transversal gates for surface codes).
Therefore, the main circuit of our scheme requires $7$ code cycles, which is $\sim d$ times faster than $8.5 d$.
Note that the $S$~state preparation for the $S$ gate has a time overhead, but as discussed in \cref{sec:hardware_analysis}, this can be performed in parallel with the preparation of noisy input $T$~states and thus does not contribute to the effective time needed. 

As noted prior, the spacetime overhead we quote for Ref.~\cite{gidneyEfficientMagicState2019} is for the most optimistic case, in practice the spacetime saving brought by our scheme can be even larger.
For our circuit, all noisy $\ket{T}$ states can be prepared in-place.
On the other hand, depending on the target CCZ error rate, it may not be possible to do in-place $\ket{T}$ states preparation for the scheme in Ref.~\cite{gidneyEfficientMagicState2019}. We need to include additional space for routing input $T$~states on top of the $18$ logical qubits, which can expand the total space requirement to $72$ logical qubits, highlighting further space savings since our circuits require 12 logical qubits at most.
Note that routing also enables a modest decrease of time overheads in Ref.~\cite{gidneyEfficientMagicState2019}, from $8.5d$ to $5.5d$ code cycles.

Our comparison above has not taken into account the differences between the underlying hardware and architecture, which ultimately determine the practical overheads.
In \cref{sec:hardware_analysis}, we compare the spacetime cost of both schemes on the same hardware -- the silicon spin qubit platform -- but in two different architectures, one shuttling-based and the other planar.
We show that even when taking into account the shuttling time needed to enable the transversal CNOTs, and the shorter code cycle time of the planar architecture, our CCZ synthillation scheme can still lead to a factor of $5$ to $7.5$ reduction in the spacetime volume of the CCZ factory (including the preparation of the noisy input $T$ states via cultivation) in the practical regime.

Next, let us consider the performance of this scheme in terms of the error rate of the output CCZ state.
Since the synthillation circuits are run at the logical level, the Clifford operations can often be implemented fault-tolerantly. Therefore, these Clifford operations are often taken to be noiseless in theoretical error analyses, the only noise source being the non-Clifford $T$ gate with rate of error $p_T$.
Under such assumptions, the output CCZ error rate from our scheme is equivalent to Ref.~\cite{gidneyEfficientMagicState2019} at lowest order $\sim28 p_T^2$.
However, in practice, the logical qubits still have non-negligible noise alongside the noisy $T$~gates due to finite distances.
We call these errors that occur due to finite distances \textit{memory errors} and for each code cycle we associate a memory error rate of $p_L$ per logical qubit.
For example, a common approximation for the surface code memory error rate is $p_L = 0.1(100p)^{(d+1)/2}$~\cite{fowlerLowOverheadQuantum2019}, for physical error rate~$p$ using a minimum weight perfect matching decoder.

Since memory errors occur per code cycle per logical qubit, for a \text{CCZ} synthillation circuit that takes $n_L$ logical qubits and $m$ code cycles we expect the resultant error rate of output \text{CCZ} states to be approximately 
\begin{align}\label{eqn:ccz_err_rate}
    p_{\textsc{ccz}} \approx n_L m p_L + 28 p_T^2.
\end{align}
Due to the smaller number of logical qubits ($n_L=8$) and the much shorter runtime ($m=7$) required by our scheme compared to Ref.~\cite{gidneyEfficientMagicState2019}, states suffer from far fewer memory errors.
This means that we can reach a lower error floor for $p_{\textsc{ccz}}$ while also reducing the contribution of input $T$~state errors $p_T$.

The performance of our scheme is shown in \cref{fig:err-rates}, using a depolarising noise model of rate $p_L$ for memory errors and $Z$~errors at rate $p_T \gg p_L$ on the input $T$~states.
$Z$~errors are used for initialised $\ket{T}$ because initial depolarising noise is identical to $Z$, and other kinds of Pauli noise can effectively be viewed as phase errors as discussed in~\cref{sec:T_noise}.
From this model we can refine the rough estimate of \cref{eqn:ccz_err_rate} to $(\frac{61}{6}+t_{\text{s}})p_L+28p_T^2$. Here $t_\text{s}$ is the number of code cycles required in the application of adaptive~$S$, including decoding time, and may be as low as 2 using the methods in \cref{sec:hardware_analysis}. Due to the problem of exponential slowdown~\cite{terhalQuantumErrorCorrection2015}, ultimately the contribution of decoding time to $t_\text{s}$ should not be large for a practical device. 

In our scheme, memory errors on auxiliary qubits 5-8 do not contribute to the $\ket{CCZ}$ infidelity at order~$p_L$, thus the effective value of $n_L$ is $4$ rather than $8$.
The remaining difference from \cref{eqn:ccz_err_rate} is mostly due to post-selection on qubit 4, which detects some memory errors in addition to the $T$ state errors it is designed to detect. It is particularly worth noting that the $Z$ errors that accumulate during the adaptive $S$ step are caught.
It is $X$~and $Y$~errors that contribute to the $t_\text{s}$~error factor from this step, both of which result in correlated phase errors.
After post-selection, these errors culminate as the~$t_\text{s}p_L$ error term. 
Furthermore, memory errors that occur on the final idling $\ket{CCZ}$ state contribute to infidelity at a rate $\frac{11}{12}p_L$, totalling to $2.75p_L$ over all 3 qubits.
Therefore, the circuit is comparatively tolerant to errors occurring at this intermediate stage.

We opt to perform $S$ gates via half-distance rotations~\cite{gidneyEfficientMagicState2019}, as discussed in~\cref{sec:hardware_analysis}. The contribution of the error from half-distance rotations can effectively be incorporated into the input error rate $p_T$ of the $T$~states involved in gate teleportation, such that the fidelity of input $S$~states should be optimised in conjunction with $T$~state fidelities.
In the case of codes with a transversal~$S$, such as the colour codes, the overheads of our scheme could be reduced further. However, the memory error rates of colour codes are comparatively higher due to the problem of decoding, which meant that they were avoided for magic state cultivation~\cite{gidney2024magicstatecultivationgrowing} even though colour codes would otherwise be a more natural fit for the later stages.

\begin{figure}[b]
    \centering
    \includegraphics{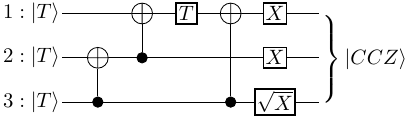}
    \caption{\label{fig:ccz-synth} Circuit for synthesising $\ket{CCZ}$ from 4 $T$ gates. Considering rate $p_L$ memory errors occurring at each CNOT and $T$ timestep, the first order contribution to $\ket{CCZ}$ infidelity is $\frac{157}{12}p_L$. If we treat idling logical qubits as ideal, then this reduces to $\frac{103}{12}p_L$.
    Note that this is not including time for adaptive $S$ feedback, and that any $\ket{T}$ idling time will directly contribute to the first order error rate, unlike the synthillation scheme of \cref{fig:impl-circ}.}
\end{figure}

Note that we focus on the regime of $p_T \gg p_L$ because when the $T$ gate error rate is comparable to the fault-tolerant Clifford operations then no synthillation is necessary and direct synthesis will produce \text{CCZ} states of similar fidelity.
That being said, the error floor of the synthesis circuit described in~\cref{fig:ccz-synth} is comparable to the synthillation circuit, meaning it may be preferable to synthillate \text{CCZ} states in most cases if a sufficient number of \text{CCZ} states are being prepared in parallel.

We considered other stochastic noise models, but the effect of logical noise bias~\cite{fazio2024logicalnoisebiasmagic} on memory errors and input states had a negligible impact on the performance and so the simplest stochastic model is presented here.
Most memory errors contribute to the infidelity of $\ket{CCZ}$ at first order, regardless of the specific type of error, hence the weak dependence.

\section{Discussion}\label{sec:discussion}
In this paper, by exploiting the transversal CNOT available in many hardware platforms, we are able to construct fault-tolerant circuits for generating distilled $\ket{CCZ}$, $\ket{CS}$ and $\ket{T}$ states with very low spacetime overhead (\cref{fig:CCZCircFinal,fig:CS-rearranged,fig:OurTgate}).
This is done through an algorithm that compiles a series of $\pi/8$-phase rotations into a circuit with low CNOT depth and minimal $T$-depth, assuming free SWAPs given by the symmetry of the circuit or added connectivity of the hardware. This algorithm can be easily generalised to arbitrary phase rotations and also rotations in other bases, thus can be useful for compiling circuits in other contexts beyond magic state preparation. It will also be interesting to generalise this algorithm to rotations of mixed basis using e.g. ideas in grouping commuting Pauli operators~\cite{gokhale2020optimization}. 

The CCZ circuit we obtain requires at most $12$ logical qubits and a minimum time overhead of $7$ code cycles. On the other hand, circuits of the same $T$ error suppression power carried out using lattice surgery with planar connectivity will require $18$ logical qubits and $8.5d$ code cycles~\cite{gidneyEfficientMagicState2019}, which is a factor of $\sim 2d$ larger in the spacetime overhead.
This shows that the additional connectivity enabling transversal CNOTs can indeed greatly reduce the spacetime overhead for magic state circuits. Note that this does not require full 3D connectivity, but only requires a thin slice of 3D structure with shallow height since the number of code patches needed in our circuits is small. The architecture to implement such a quasi-3D connectivity in 2D hardware using only local shuttling has been proposed in \cite{caiLoopedPipelinesEnabling2023}.
Besides being able to achieve the same $T$ error suppression power as its lattice surgery counterpart, our circuit will also accumulate fewer code cycle errors due to the much shorter time required. 

As mentioned in the last section, our circuits can be a key stage in magic state factories, but a previous stage is needed to prepare the incoming $T$ states~\cite{itogawa2024ZeroLevelDistillation, chamberlandVeryLowOverhead2020, gidney2024magicstatecultivationgrowing}. It will also be interesting to consider whether the availability of transversal CNOTs, along with the methods used in this paper, can be used to further optimise $T$ state preparation schemes. 

\section*{Acknowledgements}
The authors would like to thank Craig Gidney and Arthur Pesah for valuable discussions.
NF thanks his advisors Robin Harper and Stephen Bartlett for supporting this work and acknowledges support from the Australian Research Council via the Centre of Excellence in Engineered Quantum Systems (EQUS) project number CE170100009, and by the ARO under Grant Number: W911NF-21-1-0007.
ZC acknowledges support from the EPSRC QCS Hub EP/T001062/1, EPSRC projects Robust and Reliable Quantum Computing (RoaRQ, EP/W032635/1), Software Enabling Early Quantum Advantage (SEEQA, EP/Y004655/1), the Junior Research Fellowship from St John’s College, Oxford, and the EPSRC quantum technologies career acceleration fellowship (UKRI1226).
MW acknowledges support from the EPSRC [grant numbers EP/S005021/1,EP/W032635/1].

\appendix
\section{Circuit Compilation Tricks}\label{sec:circ_compilation_tricks}
Here is a list of circuit compilation tricks used to manipulate the circuits of the main text.

\paragraph{Off-loading $Z$ Rotations} $Z$ rotations (including $T$ gates) can be implemented in an out-of-place manner.
\begin{center}
    \includegraphics{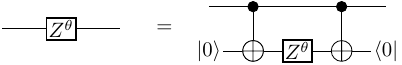}
\end{center}

\paragraph{Error propagation of \text{CCZ}}
\text{CCZ} is a gate symmetric under the permutation of all three qubits. An $X$ error on any one of the qubit will propagate to CZ on the other two qubits. 
\begin{center}
    \includegraphics{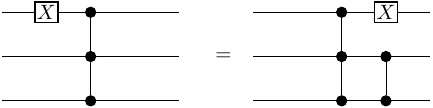}
\end{center}
\paragraph{Commuting \text{CNOT}s}
An example of commuting one \text{CNOT} through another is shown here.
\begin{center}
    \includegraphics{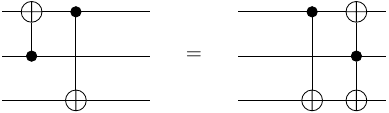}
\end{center}
We are essentially commuting $X$ in one of the \text{CNOT}s through the other \text{CNOT} to become $X \otimes X$ with both $X$s controlled on the same qubit as the original \text{CNOT} before being permuted through.

When commuting a \text{CNOT} through another \text{CNOT} that acts on the same qubits but in reverse direction, the \text{CNOT} becomes a \text{SWAP}.
\begin{center}
    \includegraphics{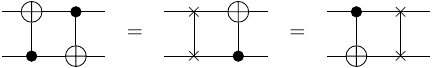}
\end{center}
\paragraph{\text{CNOT} on $\ket{+}$}
\text{CNOT}s act trivially on tensor products of $\ket{+}$ since $X$ gate acts trivially on $\ket{+}$. 

\paragraph{\text{CNOT} and \text{CCZ}}
As we try to commute \text{CNOT} through \text{CCZ}, when the control qubit is in $\ket{0}$, nothing will happen, while when the control qubit is in $\ket{1}$, the $X$ in the \text{CNOT} will propagate through the \text{CZ} in \text{CCZ} and become $X \otimes Z$. Correspondingly, it means that the \text{CNOT} will become a \text{CNOT} and a \text{CZ} when permute through \text{CCZ}. We can further permute \text{CZ} back through the \text{CCZ} while adding an $X$ gate and using rule (b) above.
\begin{center}
    \includegraphics{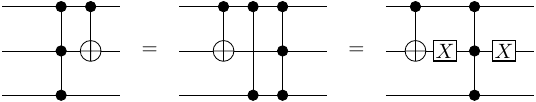}
\end{center}
Similarly, when a \text{CNOT} acts on a \text{CCZ} state we have:
\begin{center}
    \includegraphics{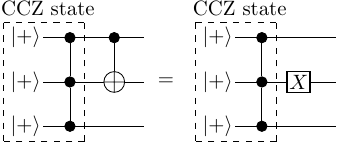}
\end{center}

\paragraph{\text{CNOT} and \text{CCZ} (single overlapping qubit)}
When the overlapping qubit is the control of \text{CNOT}, then \text{CNOT} trivially commutes through \text{CCZ}.
Otherwise, when the overlapping qubit is the target of \text{CNOT}, commuting \text{CNOT} through \text{CCZ} sends \text{CCZ} to \text{CC}($Z\otimes Z$), where the new controlled $Z$ is applied to the control qubit of \text{CNOT}.

\begin{center}
    \includegraphics{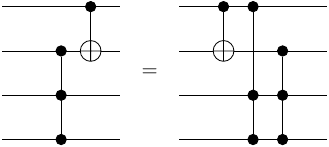}
\end{center}

\paragraph{Transforming between $T$ and $T^\dagger$}
$T$ and $T^\dagger$ can be transformed into one another by conjugating with $X$ gates. We likewise have $X \ket{T} = e^{i\pi/8} \ket{T^\dagger}$.

\paragraph{\text{CNOT} and CS} Commuting CS through CNOT simply flips the phase of CS to $\text{SI}\cdot \text{C}\mathrm{S}^\dagger$, which is the same as commuting CS though an $X$ gate on the target qubits. Hence, we have the following identity for CNOT acting on the \text{CS} state:
\begin{center}
    \includegraphics{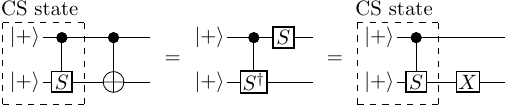}
\end{center}

\section{Faulty $\ket{T}$ states}\label{sec:T_noise}
When talking about the fault tolerance of the distillation and synthillation schemes considered in this work we start by adopting a common assumption that $\ket{T}$ states are the faulty elements in our circuits and other Clifford components (mainly \text{CNOT}s) are implemented ideally. 
In this regard we have used $T$ gates and $\ket{T}$ states interchangeably throughout this work since we can implement $T$ with $\ket{T}$ by means of an adaptive gadget.

When $T$ gates are implemented through such a gadget the associated noise channel typically takes the form 
\[p_I\rho + p_{S} S\rho S^\dag+ p_{S^\dag}S^\dag\rho S + p_{Z} Z\rho Z^\dag \]

Exactly how the channel error rates $p_S$, $p_Z$ and $p_{S^\dag}$ relate to the Pauli error rates of the original faulty $T$~gate or state will depend on the specific gadget, but for many known gadgets the implemented $T$ gates have noise of this form.

In our schemes however, not all $\ket{T}$ states are used for injection; 
the data qubits are prepared in $\ket{T}$, meaning that $X$, $Y$ and $Z$ errors are introduced to the data qubits without first being mediated by a gadget.
This does not impact the fault tolerance of these schemes because
$ X\ket{T} = e^{i\pi/8}\ket{T^\dag}= e^{i\pi/8}S^\dag\ket{T}$ and similarly $Y\ket{T} = e^{-i\pi/8}Z\ket{T^\dag} = e^{-i\pi/8}S\ket{T}$.
Therefore, the noise model of the $T$ states is the same as from the $T$ gates of our scheme, though there may be differences in the specific error rates from the gadget used. 
For $\ket{T}$ with Pauli error rates $p'_X$, $p'_Y$ and $p'_Z$ we have $p_{S^\dag} = p'_X$, $p_{S} = p'_Y$ and $p_Z=p'_Z$.

As a final note: $S\rho S^\dag + S^\dag \rho S = \rho +Z\rho Z^\dag$, meaning that many sources of $S$ and $S^\dag$ noise can be collected and re-expressed as $Z$~noise when $p_S = p_{S^\dag}$.

\section{Physical Implementation} \label{sec:hardware_analysis}
In the main text, we have not discussed the implementation cost of transversal CNOTs. Here we will take the explicit example of semi-conductor spin qubits implemented in the looped pipeline architecture~\cite{caiLoopedPipelinesEnabling2023} to illustrate the significant practical improvement brought by our distillation circuit. 

\begin{figure}[htbp]
    \centering
    \subfloat[A five-qubit array in linear shuttling tracks. \label{fig:linear_shuttle}]{\includegraphics[width = 0.45\textwidth]{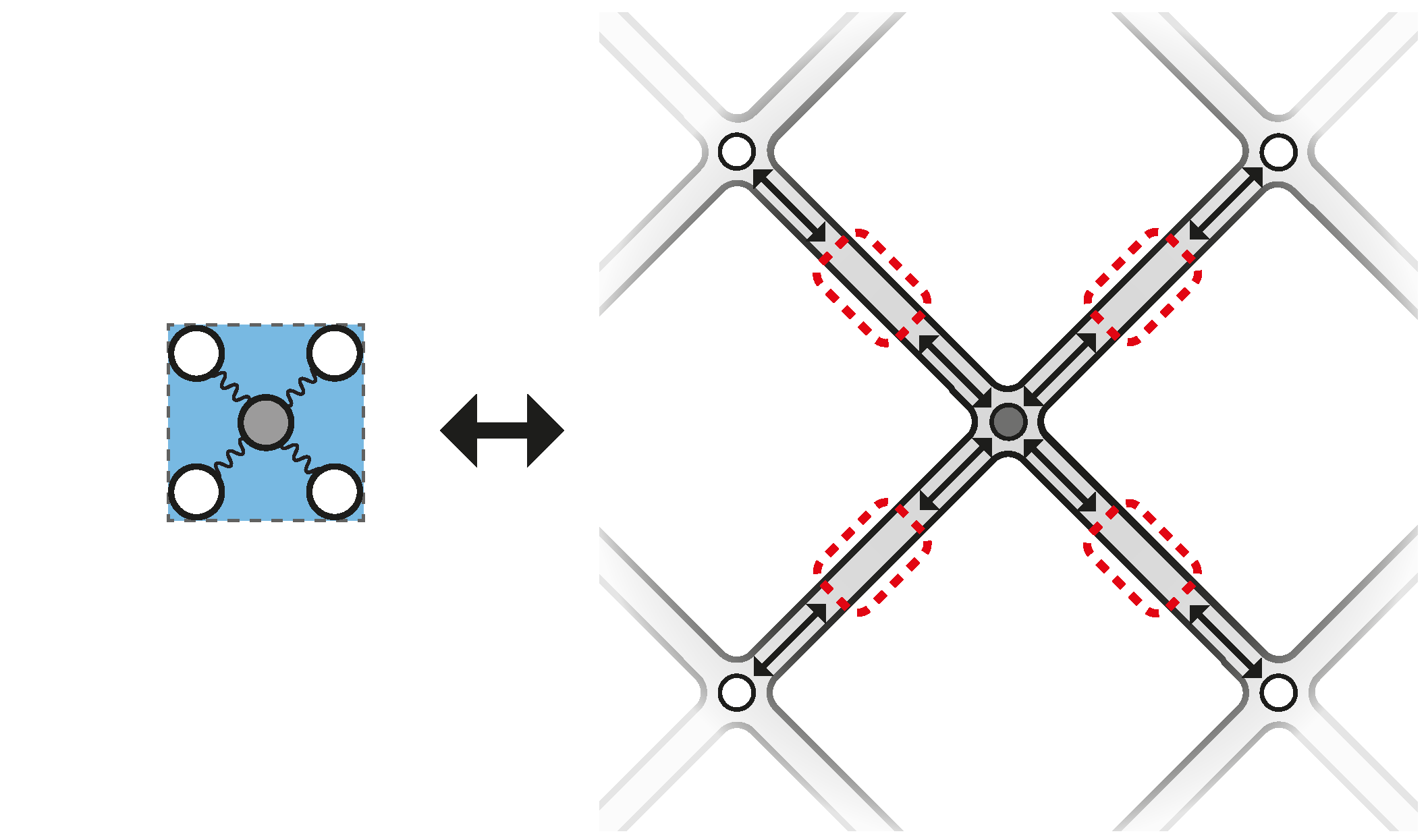}}\\
    \subfloat[A five-qubit array in shuttling loops. \label{fig:loop_shuttle}]{\includegraphics[width = 0.45\textwidth]{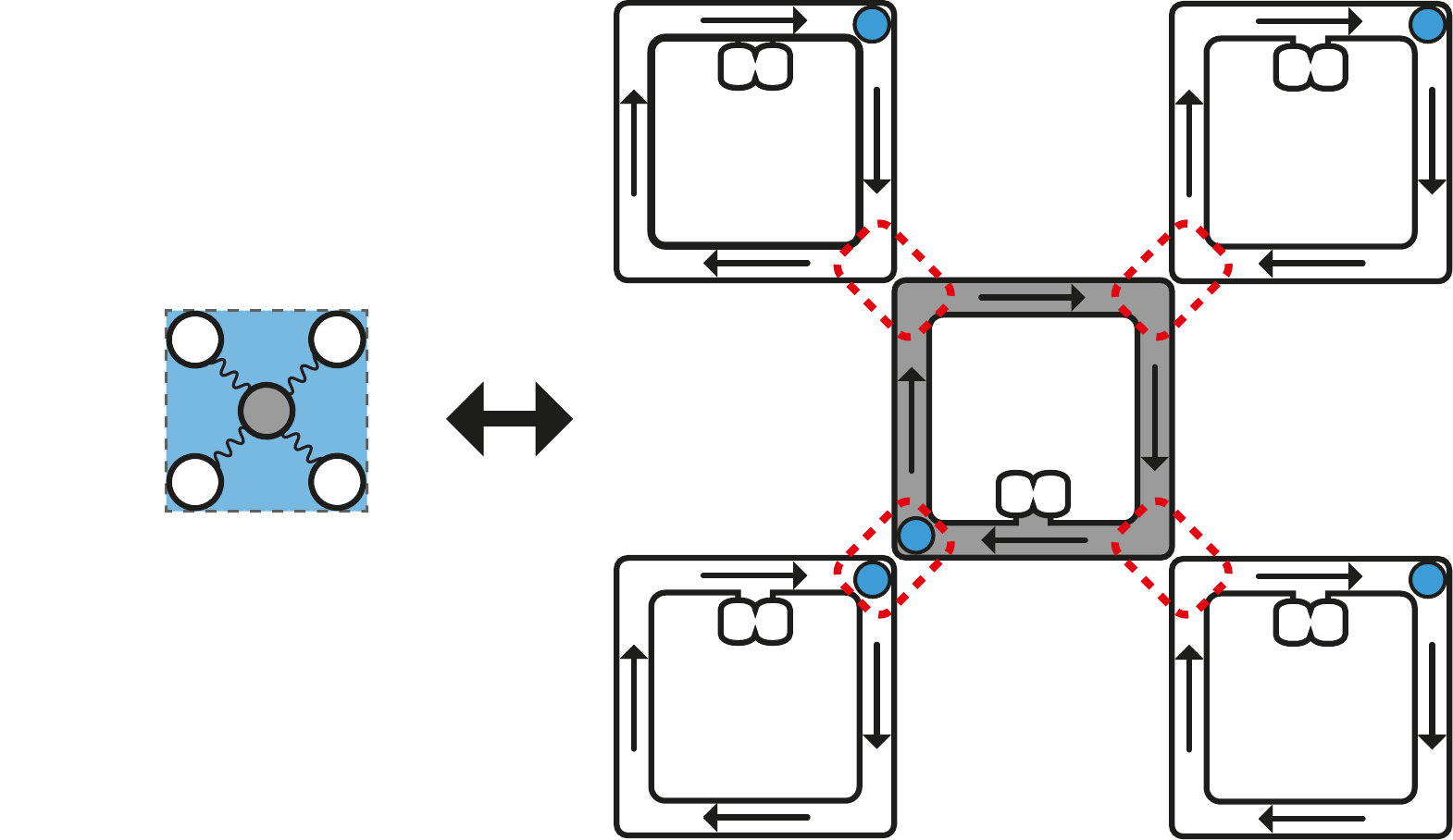}}\\
    \subfloat[Four five-qubit arrays in shuttling loops using pipelining.\label{fig:loop_pipeline}]{\includegraphics[width = 0.45\textwidth]{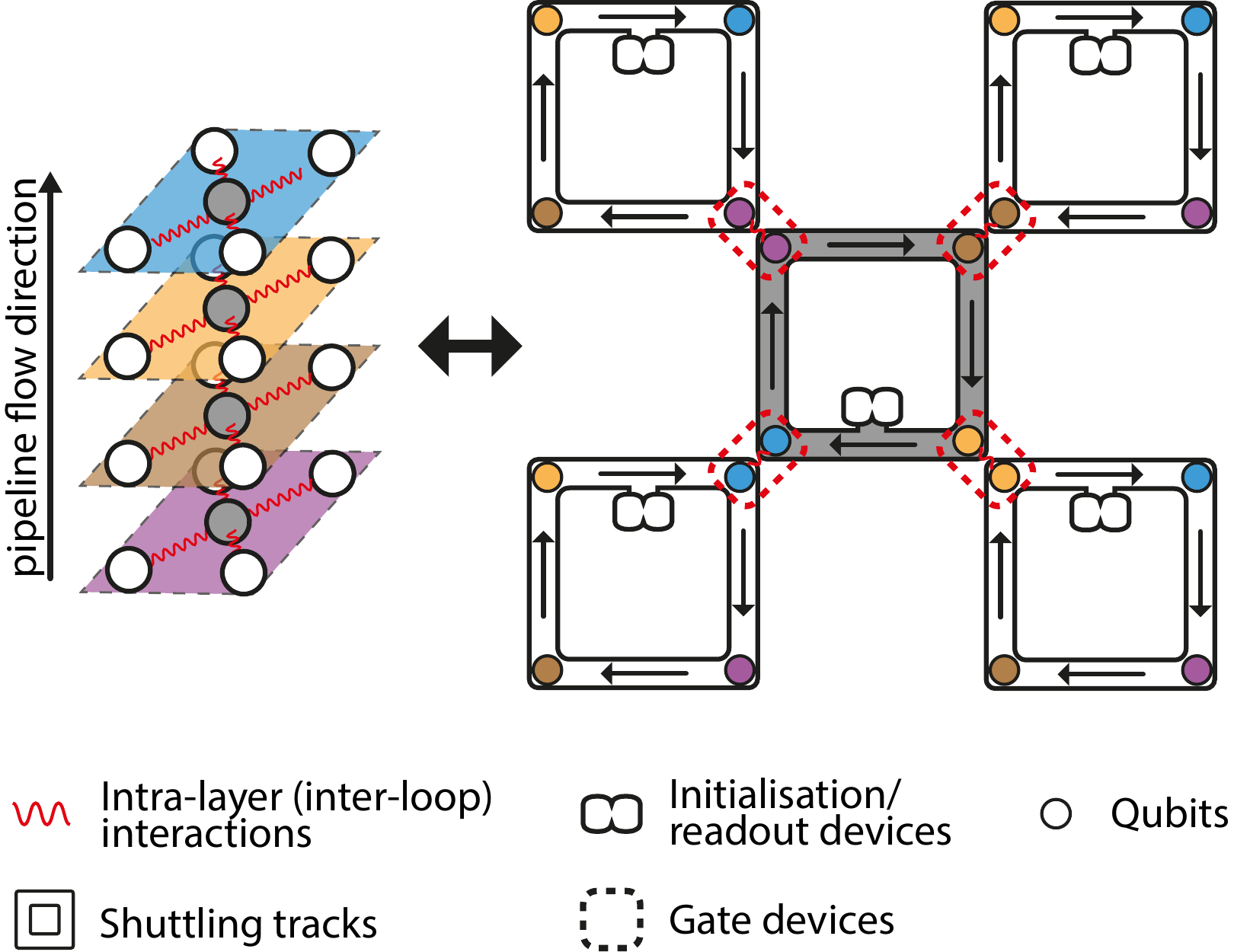}}
    \caption{Five-qubit arrays implemented using different shuttling architectures. Multiple qubits can be stored on the same loop through pipelining, which corresponds to multiple layers of five-qubit arrays.}
    \label{fig:array_pipeline_both}
\end{figure}

The ``looped pipeline'' is an architecture that utilises localised shuttling to enable efficient transversal operations between layers of qubits.
In semi-conductor platforms, linear shuttling tracks as shown in \cref{fig:linear_shuttle} can be employed to reduce crosstalk and provide space for classical controls.
As shown in \cref{fig:loop_shuttle}, the linear shuttling tracks can be replaced using shuttling loops, and qubits in the qubit array can be connected by synchronising the movements in different loops. With this new layout, we can actually fit another qubit into each loop that follows the exact same trajectory as the first qubit in the loop, but just a step behind. Hence, these new qubits in the loops will achieve the exact same qubit connectivity as the first set of qubits, creating another qubit array with no additional space overhead. In this way, by fitting $K$ qubits per loop, we can implement a stack of $K$ independent 2D qubit arrays in this ``looped pipeline'' architecture as shown in \cref{fig:loop_pipeline}.

\begin{figure}[htbp]
    \centering
    \includegraphics[width = 0.45\textwidth]{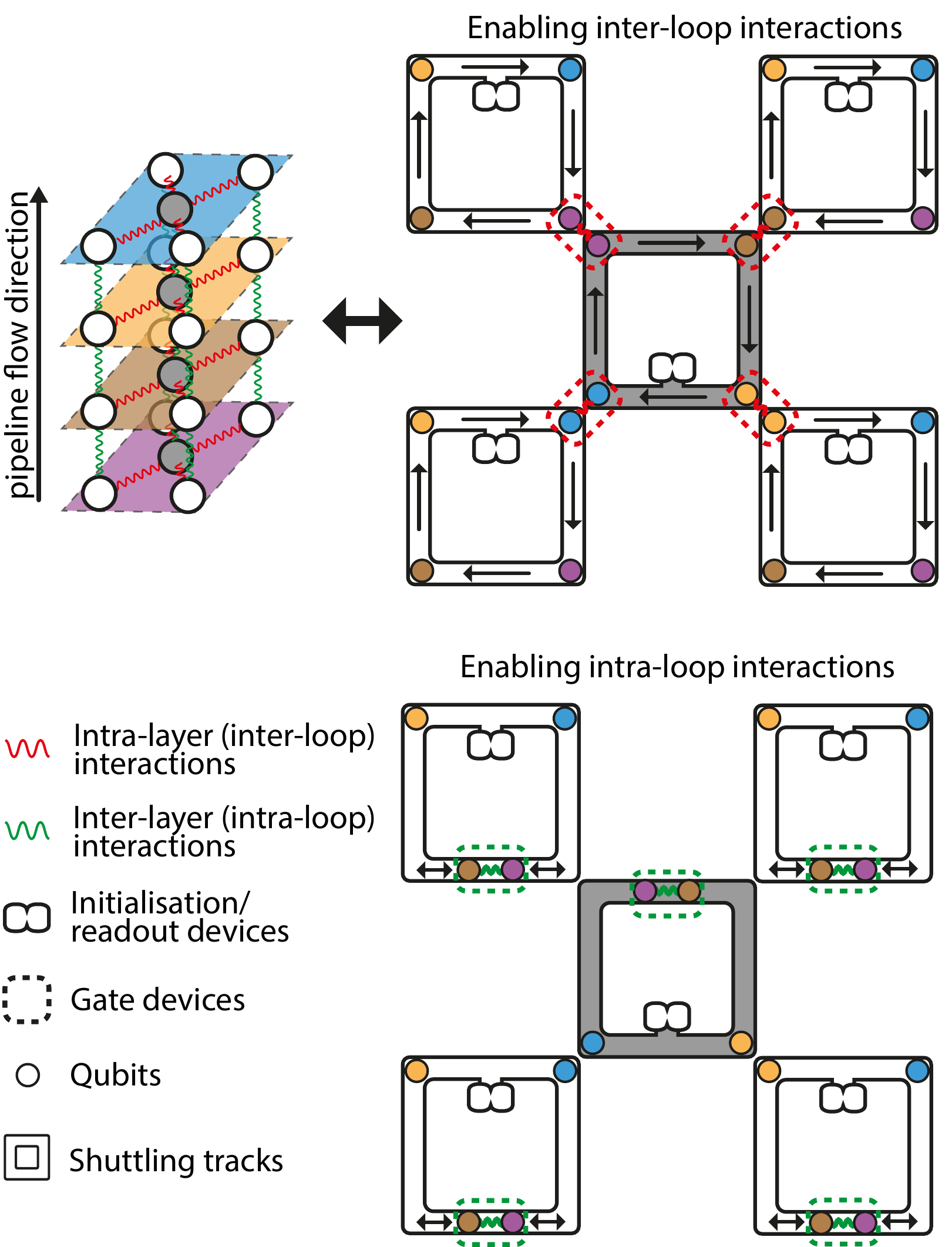}
    \caption{Interactions within layers are enabled by inter-loop (red) interactions while transversal interaction in between layers are enabled by intra-loop (green) interactions.
    }
    \label{fig:array_pipeline_intra}
\end{figure}

Now if we allow for interaction between different qubits in the same loop, we are essentially connecting qubits in different layers of qubit arrays, which enables transversal operations between different qubit arrays in the stack, as shown in \cref{fig:array_pipeline_intra}.

The number of qubits we can fit into each loop depends on the type of operations we want to perform on the qubits as they go around, the rate-limiting step, and the amount of buffering time we are willing to put in. For semi-conductor spin qubit platforms, as outlined in Ref.~\cite{caiLoopedPipelinesEnabling2023}, a reasonable loop size is with a side length of $\sim 7 \mu m$ and the shuttling speed is around $25 m s^{-1}$. Thus, the time needed to shuttle around the full loop is around $\tau_{\mathrm{sh}} \sim 1 \mu s$. The time needed on the platform for single-qubit gates, two-qubit gates, and measurements is around $\tau_{\mathrm{1qb}} \sim 25 ns$, $\tau_{\mathrm{2qb}} \sim  0.1 \mu s$, and $\tau_{\mathrm{meas}} \sim  1 \mu s$, respectively~\cite{caiLoopedPipelinesEnabling2023}.
As outlined in Table III of Ref.~\cite{caiLoopedPipelinesEnabling2023}, the code cycle time for the surface code on this platform is $\sim 3.85 \mu s$. We can fit $5$ qubits in each loop with a slightly increased code cycle time of $\sim 4.8 \mu s$. Measurement is the rate-limiting step, thus by having 4 measurement devices per loop, we can fit up to $14$ qubits per loop while keeping the same code cycle time of $\sim 3.85 \mu s$ (see Appendix E of Ref.~\cite{caiLoopedPipelinesEnabling2023}). The number of measurement devices needed can be further reduced with improvements in readout technologies.

\begin{figure}[htbp]
    \centering
    \includegraphics[width = 0.45\textwidth]{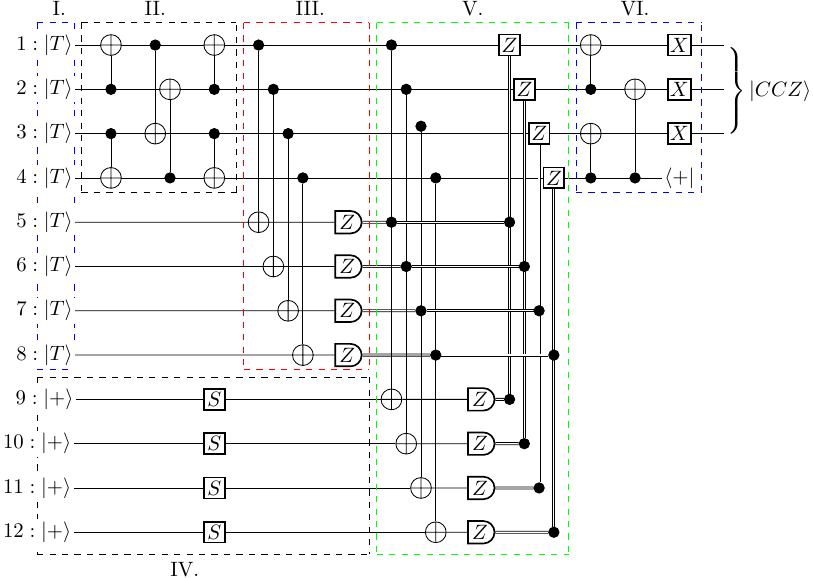}
    \caption{Full CCZ synthillation circuit with $S$ correction.}
    \label{fig:CCZIncludeS}
\end{figure}

Now working with $12$ qubits per loop, i.e. a stack of $12$ surface code layers, let us consider the implementation costs of the CCZ distillation circuit in \cref{fig:CCZIncludeS}. By adding bypasses (for qubits overtaking one another) and holding space in the shuttling loops, different qubit layers can be swapped around by physically exchanging the qubits using shuttling. With $12$ qubits per loop, the time needed for the neighbouring qubits to swap places using shuttling is $\tau_{\mathrm{sw}} = \tau_{\mathrm{sh}}/12 \sim 0.1 \mu s$. Using the bypasses and holding space, the time needed to rearrange the order of all qubits in the loop will roughly be the same as the time needed for the qubits that travel the longest distance in this rearrangement, which is roughly upper bounded by the time needed for a qubit to travel around the loop. Without zooming into the details of the layout of the holding spaces and bypasses, we will assume this is the time needed for the rearrangement of qubits $\tau_{\mathrm{perm}} \sim \tau_{\mathrm{sh}} \sim 1 \mu s$.

We now show how the CCZ synthillation circuit of \cref{fig:CCZIncludeS} is implemented using this architecture.
The spacetime complexity for each time step of the circuit is as follows:
\begin{enumerate}[label=\Roman*.]
    \item Incoming noisy $T$ states can be prepared via injection, cultivation or distillation depending on the noise level we want to achieve. For cultivation, as mentioned in \cite{gidneyHowFactor20482025}, the spacetime volume per cultivated $T$ state of error rate $10^{-7}$ is $30000$ physical qubits $\cdot$ code cycles. Here we have $8$ logical patches used for $T$ cultivation, each contains $2d^2$ physical qubits, and we need $8$ $T$ states at the end. So the time needed is $30000/(8\times 2d^2) \times 8 = 15000/d^2$. 
    \item One round of transversal CNOTs requires $\tau_{\mathrm{2qb}} \sim 0.1 \mu s$. We need to swap logical layer 2 and 3 to perform the next layer of CNOTs. The time needed for the swap along with the CNOTs $\tau_{\mathrm{sw}} + \tau_{\mathrm{2qb}} \sim 0.2 \mu s$. The same operations are needed again for the next layer of CNOT and thus are repeated. In total, the time cost of this step is $0.1 + 0.2\times2 = 0.5 \mu s$, which is negligible compared to our code cycle time of $3.85 \mu s$.
    \item Complete rearrangement of qubit order, one round of transversal CNOTs and then one round of measurement, requires $\tau_{\mathrm{perm}} + \tau_{\mathrm{2qb}} + \tau_{\mathrm{meas}} \sim 2.1 \mu s$. 
    \item Here $S$ states are prepared using half-distance rotation~\cite{brownPokingHolesCutting2017}, which takes $d$ code cycles. Half-distance $S$ can be used here because the additional errors it incurs are typically smaller than the errors of the input noisy $T$ and thus is not a dominating factor. 
    \item Re-arranging layers 1 to 4 and the four $\ket{S}$ state layers so that $\ket{S}$ can be readily teleported into layers 1 to~4 takes $\tau_{\mathrm{perm}} \sim 1 \mu s$. Teleporting the $S$ gate requires one round of two-qubit gates, one round of measurement, and one round of single-qubit gates. Thus in total we need $\tau_{\mathrm{perm}} + \tau_{\mathrm{2qb}} + \tau_{\mathrm{meas}} + \tau_{\mathrm{1qb}} \approx 2.1 \mu s$.
    \item One round of re-arrangement enables the connectivity for the CNOTs in the next two steps, which takes $\tau_{\mathrm{perm}} \sim 1 \mu s$. Combining with the two rounds of transversal CNOTs and the single round of transversal $X$ afterwards, the time needed is $\tau_{\mathrm{perm}} + 2\tau_{\mathrm{2qb}} + \tau_{\mathrm{1qb}} = 1.2 \mu s$.
\end{enumerate}
We see that $S$ state preparation can be run in parallel with part I, II and III, and the time needed simply depends on which part is rate-limiting, i.e. $\text{max}(15000/d^2,\ d)$. They are roughly the same when $d = 25$.
For larger distances, the cultivation will be faster due to the larger space available, thus, the $S$~state preparation will be the rate-limiting step. For smaller distances, the cultivation will be slower due to the smaller space available and thus is rate-limiting. The time needed for the transversal gates is negligible even with shuttling as detailed above, but we will add one additional round of stabiliser checks after each round of transversal gates. Since Pauli gates can be implemented through classical updates, there are $7$ rounds of transversal gates that we need to explicitly implement. Hence, this will give a total time of $T_{\textsc{ccz}} \sim (\text{max}(15000/d^2,\ d) + 7)$ code cycles $\sim (\text{max}(57750/d^2,\ 3.85d) + 27) \mu s$. The spacetime volume is $V_{\textsc{ccz}} \sim T_{\textsc{ccz}} \times 12 \times 2 d^2 = (\text{max}(1.4\times10^{6},\ 92d^3) + 648 d^2) \text{qubit} \cdot \mu s$.

In a semi-conductor spin qubit architecture with all its qubits densely packed together (thus no shuttling is used), its code code cycle time can be roughly estimated as the looped pipeline code cycle time of $3.85 \mu s$ minus the time needed for shuttling in each code cycle, which is $1 \mu s$, therefore $\tau_{cyc}' = 2.85 \mu s$.
The CCZ synthillation scheme using lattice surgery of \cite{gidneyEfficientMagicState2019} requires $3 \times 6 = 18$ logical patches.
The whole circuit takes $8.5d$ code cycles, and since in this case we must perform cultivation in-place on these logical patches, this runtime cannot be easily reduced by staggering adjacent rounds as mentioned in \cite{gidneyEfficientMagicState2019}.
The cultivation stage will require $30000/(18\times 2\times d^2) \times 8 = 6667/d^2$, so the total time needed is $T_{\textsc{ccz}}' \sim (6667/d^2 + 8.5 d ) \times 2.85 \mu s \approx (19000/d^2 + 24d)\mu s$. The spacetime volume is $V_{\textsc{ccz}}' \sim T_{\textsc{ccz}}' \times 18 \times 2 d^2 = (7 \times 10^{5} + 864d^3) \text{qubit} \cdot \mu s$.

\begin{figure}[htbp]
    \centering
    \includegraphics[width = 0.45\textwidth]{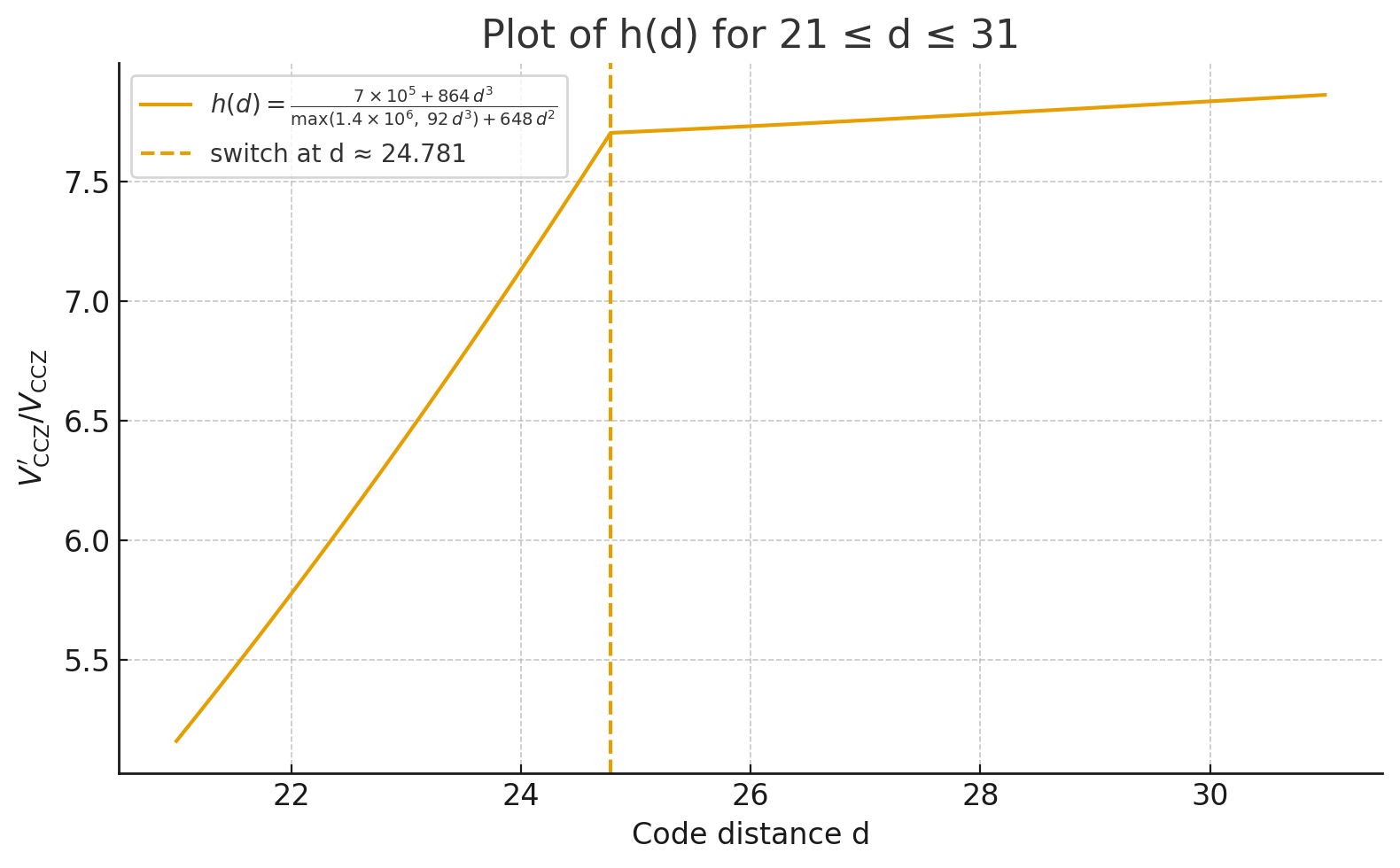}
    \caption{Factor of spacetime saving using our circuit, taking into account of the shuttling time required.
    }
    \label{fig:CCZ_SpaceTime_saving}
\end{figure}

Therefore, the factor of space-time volume saving for our circuit over the conventional approach, even taking into account the shuttling time, is $V_{\textsc{ccz}}'/V_{\textsc{ccz}} = \frac{7 \times 10^{5} + 864d^3}{\text{max}(1.4\times10^{6},\ 92d^3) + 648 d^2}$. However, note that this figure is calculated using a cultivation error rate of $10^{-7}$, which will output a CCZ error rate of $\sim 10^{-13}$. Such a low error rate is only necessary when we need to work with codes of distance $21$ and above. In \cref{fig:CCZ_SpaceTime_saving}, we have plotted $V_{\textsc{ccz}}'/V_{\textsc{ccz}}$ for $21 \leq d \leq 31$. We see that our CCZ factory provides a spacetime saving in this entire regime, by a factor ranging between 5 and 7.5. 

Now if we disregard the time required for the zero-level cultivation and focus entirely on the time saving in the distillation circuit, for our circuit the time needed is $(\text{max}(0,\ d-15000/d^2) + 7)$ code cycles, which for the practical regime of $d \lesssim 25$, is roughly $7$ and thus independent of $d$. On the other hand, as mentioned prior, the scheme in using lattice surgery in \cite{gidneyEfficientMagicState2019,gidney2024magicstatecultivationgrowing} requires $8.5d$ code cycles, which is much larger.

%


\begin{thebibliography}{30}%
\makeatletter
\providecommand \@ifxundefined [1]{%
 \@ifx{#1\undefined}
}%
\providecommand \@ifnum [1]{%
 \ifnum #1\expandafter \@firstoftwo
 \else \expandafter \@secondoftwo
 \fi
}%
\providecommand \@ifx [1]{%
 \ifx #1\expandafter \@firstoftwo
 \else \expandafter \@secondoftwo
 \fi
}%
\providecommand \natexlab [1]{#1}%
\providecommand \enquote  [1]{``#1''}%
\providecommand \bibnamefont  [1]{#1}%
\providecommand \bibfnamefont [1]{#1}%
\providecommand \citenamefont [1]{#1}%
\providecommand \href@noop [0]{\@secondoftwo}%
\providecommand \href [0]{\begingroup \@sanitize@url \@href}%
\providecommand \@href[1]{\@@startlink{#1}\@@href}%
\providecommand \@@href[1]{\endgroup#1\@@endlink}%
\providecommand \@sanitize@url [0]{\catcode `\\12\catcode `\$12\catcode
  `\&12\catcode `\#12\catcode `\^12\catcode `\_12\catcode `\%12\relax}%
\providecommand \@@startlink[1]{}%
\providecommand \@@endlink[0]{}%
\providecommand \url  [0]{\begingroup\@sanitize@url \@url }%
\providecommand \@url [1]{\endgroup\@href {#1}{\urlprefix }}%
\providecommand \urlprefix  [0]{URL }%
\providecommand \Eprint [0]{\href }%
\providecommand \doibase [0]{https://doi.org/}%
\providecommand \selectlanguage [0]{\@gobble}%
\providecommand \bibinfo  [0]{\@secondoftwo}%
\providecommand \bibfield  [0]{\@secondoftwo}%
\providecommand \translation [1]{[#1]}%
\providecommand \BibitemOpen [0]{}%
\providecommand \bibitemStop [0]{}%
\providecommand \bibitemNoStop [0]{.\EOS\space}%
\providecommand \EOS [0]{\spacefactor3000\relax}%
\providecommand \BibitemShut  [1]{\csname bibitem#1\endcsname}%
\let\auto@bib@innerbib\@empty
\bibitem [{\citenamefont {Postler}\ \emph {et~al.}(2022)\citenamefont
  {Postler}, \citenamefont {Heu{\ss}en}, \citenamefont {Pogorelov},
  \citenamefont {Rispler}, \citenamefont {Feldker}, \citenamefont {Meth},
  \citenamefont {Marciniak}, \citenamefont {Stricker}, \citenamefont
  {Ringbauer}, \citenamefont {Blatt}, \citenamefont {Schindler}, \citenamefont
  {M{\"u}ller},\ and\ \citenamefont
  {Monz}}]{postlerDemonstrationFaulttolerantUniversal2022}%
  \BibitemOpen
  \bibfield  {author} {\bibinfo {author} {\bibfnamefont {L.}~\bibnamefont
  {Postler}}, \bibinfo {author} {\bibfnamefont {S.}~\bibnamefont {Heu{\ss}en}},
  \bibinfo {author} {\bibfnamefont {I.}~\bibnamefont {Pogorelov}}, \bibinfo
  {author} {\bibfnamefont {M.}~\bibnamefont {Rispler}}, \bibinfo {author}
  {\bibfnamefont {T.}~\bibnamefont {Feldker}}, \bibinfo {author} {\bibfnamefont
  {M.}~\bibnamefont {Meth}}, \bibinfo {author} {\bibfnamefont {C.~D.}\
  \bibnamefont {Marciniak}}, \bibinfo {author} {\bibfnamefont {R.}~\bibnamefont
  {Stricker}}, \bibinfo {author} {\bibfnamefont {M.}~\bibnamefont {Ringbauer}},
  \bibinfo {author} {\bibfnamefont {R.}~\bibnamefont {Blatt}}, \bibinfo
  {author} {\bibfnamefont {P.}~\bibnamefont {Schindler}}, \bibinfo {author}
  {\bibfnamefont {M.}~\bibnamefont {M{\"u}ller}},\ and\ \bibinfo {author}
  {\bibfnamefont {T.}~\bibnamefont {Monz}},\ }\bibfield  {title} {\bibinfo
  {title} {Demonstration of fault-tolerant universal quantum gate operations},\
  }\href {https://doi.org/10.1038/s41586-022-04721-1} {\bibfield  {journal}
  {\bibinfo  {journal} {Nature}\ }\textbf {\bibinfo {volume} {605}},\ \bibinfo
  {pages} {675} (\bibinfo {year} {2022})}\BibitemShut {NoStop}%
\bibitem [{\citenamefont {Bluvstein}\ \emph {et~al.}(2024)\citenamefont
  {Bluvstein} \emph {et~al.}}]{bluvsteinLogicalQuantumProcessor2024}%
  \BibitemOpen
  \bibfield  {author} {\bibinfo {author} {\bibfnamefont {D.}~\bibnamefont
  {Bluvstein}} \emph {et~al.},\ }\bibfield  {title} {\bibinfo {title} {Logical
  quantum processor based on reconfigurable atom arrays},\ }\href
  {https://doi.org/10.1038/s41586-023-06927-3} {\bibfield  {journal} {\bibinfo
  {journal} {Nature}\ }\textbf {\bibinfo {volume} {626}},\ \bibinfo {pages}
  {58} (\bibinfo {year} {2024})}\BibitemShut {NoStop}%
\bibitem [{\citenamefont {{Ryan-Anderson}}\ \emph {et~al.}(2024)\citenamefont
  {{Ryan-Anderson}} \emph
  {et~al.}}]{ryan-andersonHighfidelityTeleportationLogical2024}%
  \BibitemOpen
  \bibfield  {author} {\bibinfo {author} {\bibfnamefont {C.}~\bibnamefont
  {{Ryan-Anderson}}} \emph {et~al.},\ }\bibfield  {title} {\bibinfo {title}
  {High-fidelity teleportation of a logical qubit using transversal gates and
  lattice surgery},\ }\href {https://doi.org/10.1126/science.adp6016}
  {\bibfield  {journal} {\bibinfo  {journal} {Science}\ }\textbf {\bibinfo
  {volume} {385}},\ \bibinfo {pages} {1327} (\bibinfo {year}
  {2024})}\BibitemShut {NoStop}%
\bibitem [{\citenamefont {Acharya}\ \emph {et~al.}(2024)\citenamefont {Acharya}
  \emph {et~al.}}]{googleQECBelowThreshold2024}%
  \BibitemOpen
  \bibfield  {author} {\bibinfo {author} {\bibfnamefont {R.}~\bibnamefont
  {Acharya}} \emph {et~al.},\ }\bibfield  {title} {\bibinfo {title} {Quantum
  error correction below the surface code threshold},\ }\bibfield  {journal}
  {\bibinfo  {journal} {Nature}\ }\href
  {https://doi.org/10.1038/s41586-024-08449-y} {10.1038/s41586-024-08449-y}
  (\bibinfo {year} {2024})\BibitemShut {NoStop}%
\bibitem [{\citenamefont
  {Litinski}(2019{\natexlab{a}})}]{litinskiMagicStateDistillation2019}%
  \BibitemOpen
  \bibfield  {author} {\bibinfo {author} {\bibfnamefont {D.}~\bibnamefont
  {Litinski}},\ }\bibfield  {title} {\bibinfo {title} {Magic {{State
  Distillation}}: {{Not}} as {{Costly}} as {{You Think}}},\ }\href
  {https://doi.org/10.22331/q-2019-12-02-205} {\bibfield  {journal} {\bibinfo
  {journal} {Quantum}\ }\textbf {\bibinfo {volume} {3}},\ \bibinfo {pages}
  {205} (\bibinfo {year} {2019}{\natexlab{a}})}\BibitemShut {NoStop}%
\bibitem [{\citenamefont {Chamberland}\ and\ \citenamefont
  {Noh}(2020)}]{chamberlandVeryLowOverhead2020}%
  \BibitemOpen
  \bibfield  {author} {\bibinfo {author} {\bibfnamefont {C.}~\bibnamefont
  {Chamberland}}\ and\ \bibinfo {author} {\bibfnamefont {K.}~\bibnamefont
  {Noh}},\ }\bibfield  {title} {\bibinfo {title} {Very low overhead
  fault-tolerant magic state preparation using redundant ancilla encoding and
  flag qubits},\ }\href {https://doi.org/10.1038/s41534-020-00319-5} {\bibfield
   {journal} {\bibinfo  {journal} {npj Quantum Information}\ }\textbf {\bibinfo
  {volume} {6}},\ \bibinfo {pages} {91} (\bibinfo {year} {2020})}\BibitemShut
  {NoStop}%
\bibitem [{\citenamefont {Itogawa}\ \emph {et~al.}(2024)\citenamefont
  {Itogawa}, \citenamefont {Takada}, \citenamefont {Hirano},\ and\
  \citenamefont {Fujii}}]{itogawa2024ZeroLevelDistillation}%
  \BibitemOpen
  \bibfield  {author} {\bibinfo {author} {\bibfnamefont {T.}~\bibnamefont
  {Itogawa}}, \bibinfo {author} {\bibfnamefont {Y.}~\bibnamefont {Takada}},
  \bibinfo {author} {\bibfnamefont {Y.}~\bibnamefont {Hirano}},\ and\ \bibinfo
  {author} {\bibfnamefont {K.}~\bibnamefont {Fujii}},\ }\href
  {https://arxiv.org/abs/2403.03991} {\bibinfo {title} {Even more efficient
  magic state distillation by zero-level distillation}} (\bibinfo {year}
  {2024}),\ \Eprint {https://arxiv.org/abs/2403.03991} {arXiv:2403.03991
  [quant-ph]} \BibitemShut {NoStop}%
\bibitem [{\citenamefont {Gidney}\ \emph {et~al.}(2024)\citenamefont {Gidney},
  \citenamefont {Shutty},\ and\ \citenamefont
  {Jones}}]{gidney2024magicstatecultivationgrowing}%
  \BibitemOpen
  \bibfield  {author} {\bibinfo {author} {\bibfnamefont {C.}~\bibnamefont
  {Gidney}}, \bibinfo {author} {\bibfnamefont {N.}~\bibnamefont {Shutty}},\
  and\ \bibinfo {author} {\bibfnamefont {C.}~\bibnamefont {Jones}},\ }\href
  {https://arxiv.org/abs/2409.17595} {\bibinfo {title} {Magic state
  cultivation: growing t states as cheap as cnot gates}} (\bibinfo {year}
  {2024}),\ \Eprint {https://arxiv.org/abs/2409.17595} {arXiv:2409.17595
  [quant-ph]} \BibitemShut {NoStop}%
\bibitem [{\citenamefont
  {Jones}(2013)}]{jonesLowoverheadConstructionsFaulttolerant2013}%
  \BibitemOpen
  \bibfield  {author} {\bibinfo {author} {\bibfnamefont {C.}~\bibnamefont
  {Jones}},\ }\bibfield  {title} {\bibinfo {title} {Low-overhead constructions
  for the fault-tolerant {{Toffoli}} gate},\ }\href
  {https://doi.org/10.1103/PhysRevA.87.022328} {\bibfield  {journal} {\bibinfo
  {journal} {Phys. Rev. A}\ }\textbf {\bibinfo {volume} {87}},\ \bibinfo
  {pages} {022328} (\bibinfo {year} {2013})}\BibitemShut {NoStop}%
\bibitem [{\citenamefont {Amy}\ \emph {et~al.}(2014)\citenamefont {Amy},
  \citenamefont {Maslov},\ and\ \citenamefont {Mosca}}]{matroid}%
  \BibitemOpen
  \bibfield  {author} {\bibinfo {author} {\bibfnamefont {M.}~\bibnamefont
  {Amy}}, \bibinfo {author} {\bibfnamefont {D.}~\bibnamefont {Maslov}},\ and\
  \bibinfo {author} {\bibfnamefont {M.}~\bibnamefont {Mosca}},\ }\bibfield
  {title} {\bibinfo {title} {Polynomial-time t-depth optimization of clifford+t
  circuits via matroid partitioning},\ }\href
  {https://doi.org/10.1109/TCAD.2014.2341953} {\bibfield  {journal} {\bibinfo
  {journal} {IEEE Transactions on Computer-Aided Design of Integrated Circuits
  and Systems}\ }\textbf {\bibinfo {volume} {33}},\ \bibinfo {pages} {1476}
  (\bibinfo {year} {2014})}\BibitemShut {NoStop}%
\bibitem [{\citenamefont {Selinger}(2013)}]{selingerQuantumCircuitsDepth2013}%
  \BibitemOpen
  \bibfield  {author} {\bibinfo {author} {\bibfnamefont {P.}~\bibnamefont
  {Selinger}},\ }\bibfield  {title} {\bibinfo {title} {Quantum circuits of
  \${{T}}\$-depth one},\ }\href {https://doi.org/10.1103/PhysRevA.87.042302}
  {\bibfield  {journal} {\bibinfo  {journal} {Phys. Rev. A}\ }\textbf {\bibinfo
  {volume} {87}},\ \bibinfo {pages} {042302} (\bibinfo {year}
  {2013})}\BibitemShut {NoStop}%
\bibitem [{\citenamefont {Bravyi}\ and\ \citenamefont
  {Haah}(2012)}]{Bravyi2012MSDTriorthogonal}%
  \BibitemOpen
  \bibfield  {author} {\bibinfo {author} {\bibfnamefont {S.}~\bibnamefont
  {Bravyi}}\ and\ \bibinfo {author} {\bibfnamefont {J.}~\bibnamefont {Haah}},\
  }\bibfield  {title} {\bibinfo {title} {Magic-state distillation with low
  overhead},\ }\href {https://doi.org/10.1103/PhysRevA.86.052329} {\bibfield
  {journal} {\bibinfo  {journal} {Phys. Rev. A}\ }\textbf {\bibinfo {volume}
  {86}},\ \bibinfo {pages} {052329} (\bibinfo {year} {2012})}\BibitemShut
  {NoStop}%
\bibitem [{\citenamefont {Campbell}\ and\ \citenamefont
  {Howard}(2017)}]{campbell2017unified}%
  \BibitemOpen
  \bibfield  {author} {\bibinfo {author} {\bibfnamefont {E.~T.}\ \bibnamefont
  {Campbell}}\ and\ \bibinfo {author} {\bibfnamefont {M.}~\bibnamefont
  {Howard}},\ }\bibfield  {title} {\bibinfo {title} {Unified framework for
  magic state distillation and multiqubit gate synthesis with reduced resource
  cost},\ }\href {https://doi.org/10.1103/PhysRevA.95.022316} {\bibfield
  {journal} {\bibinfo  {journal} {Phys. Rev. A}\ }\textbf {\bibinfo {volume}
  {95}},\ \bibinfo {pages} {022316} (\bibinfo {year} {2017})}\BibitemShut
  {NoStop}%
\bibitem [{\citenamefont {Gidney}\ and\ \citenamefont
  {Fowler}(2019)}]{gidneyEfficientMagicState2019}%
  \BibitemOpen
  \bibfield  {author} {\bibinfo {author} {\bibfnamefont {C.}~\bibnamefont
  {Gidney}}\ and\ \bibinfo {author} {\bibfnamefont {A.~G.}\ \bibnamefont
  {Fowler}},\ }\bibfield  {title} {\bibinfo {title} {Efficient magic state
  factories with a catalyzed \${\textbar}{{CCZ}}{\textbackslash}rangle\$ to
  \$2{\textbar}{{T}}{\textbackslash}rangle\$ transformation},\ }\href
  {https://doi.org/10.22331/q-2019-04-30-135} {\bibfield  {journal} {\bibinfo
  {journal} {Quantum}\ }\textbf {\bibinfo {volume} {3}},\ \bibinfo {pages}
  {135} (\bibinfo {year} {2019})}\BibitemShut {NoStop}%
\bibitem [{\citenamefont {Rodriguez}\ \emph {et~al.}(2024)\citenamefont
  {Rodriguez} \emph {et~al.}}]{rodriguezExperimentalDemonstrationLogical2024}%
  \BibitemOpen
  \bibfield  {author} {\bibinfo {author} {\bibfnamefont {P.~S.}\ \bibnamefont
  {Rodriguez}} \emph {et~al.},\ }\bibfield  {title} {\bibinfo {title}
  {Experimental demonstration of logical magic state distillation},\ }\bibfield
   {journal} {\bibinfo  {journal} {arXiv}\ }\href
  {https://doi.org/10.48550/arXiv.2412.15165} {10.48550/arXiv.2412.15165}
  (\bibinfo {year} {2024})\BibitemShut {NoStop}%
\bibitem [{\citenamefont {Zhou}\ \emph {et~al.}(2024)\citenamefont {Zhou},
  \citenamefont {Zhao}, \citenamefont {Cain}, \citenamefont {Bluvstein},
  \citenamefont {Duckering}, \citenamefont {Hu}, \citenamefont {Wang},
  \citenamefont {Kubica},\ and\ \citenamefont
  {Lukin}}]{zhouAlgorithmicFaultTolerance2024}%
  \BibitemOpen
  \bibfield  {author} {\bibinfo {author} {\bibfnamefont {H.}~\bibnamefont
  {Zhou}}, \bibinfo {author} {\bibfnamefont {C.}~\bibnamefont {Zhao}}, \bibinfo
  {author} {\bibfnamefont {M.}~\bibnamefont {Cain}}, \bibinfo {author}
  {\bibfnamefont {D.}~\bibnamefont {Bluvstein}}, \bibinfo {author}
  {\bibfnamefont {C.}~\bibnamefont {Duckering}}, \bibinfo {author}
  {\bibfnamefont {H.-Y.}\ \bibnamefont {Hu}}, \bibinfo {author} {\bibfnamefont
  {S.-T.}\ \bibnamefont {Wang}}, \bibinfo {author} {\bibfnamefont
  {A.}~\bibnamefont {Kubica}},\ and\ \bibinfo {author} {\bibfnamefont {M.~D.}\
  \bibnamefont {Lukin}},\ }\bibfield  {title} {\bibinfo {title} {Algorithmic
  fault tolerance for fast quantum computing},\ }\bibfield  {journal} {\bibinfo
   {journal} {arXiv}\ }\href {https://doi.org/10.48550/arXiv.2406.17653}
  {10.48550/arXiv.2406.17653} (\bibinfo {year} {2024})\BibitemShut {NoStop}%
\bibitem [{\citenamefont
  {Litinski}(2019{\natexlab{b}})}]{litinskiGameSurfaceCodes2019}%
  \BibitemOpen
  \bibfield  {author} {\bibinfo {author} {\bibfnamefont {D.}~\bibnamefont
  {Litinski}},\ }\bibfield  {title} {\bibinfo {title} {A {{Game}} of {{Surface
  Codes}}: {{Large-Scale Quantum Computing}} with {{Lattice Surgery}}},\ }\href
  {https://doi.org/10.22331/q-2019-03-05-128} {\bibfield  {journal} {\bibinfo
  {journal} {Quantum}\ }\textbf {\bibinfo {volume} {3}},\ \bibinfo {pages}
  {128} (\bibinfo {year} {2019}{\natexlab{b}})}\BibitemShut {NoStop}%
\bibitem [{\citenamefont {Murphy}\ and\ \citenamefont
  {Kissinger}(2023)}]{CNOT_parity_matrix}%
  \BibitemOpen
  \bibfield  {author} {\bibinfo {author} {\bibfnamefont {E.}~\bibnamefont
  {Murphy}}\ and\ \bibinfo {author} {\bibfnamefont {A.}~\bibnamefont
  {Kissinger}},\ }\bibfield  {title} {\bibinfo {title} {Global synthesis of
  cnot circuits with holes},\ }\href {https://doi.org/10.4204/eptcs.384.5}
  {\bibfield  {journal} {\bibinfo  {journal} {Electronic Proceedings in
  Theoretical Computer Science}\ }\textbf {\bibinfo {volume} {384}},\ \bibinfo
  {pages} {75–88} (\bibinfo {year} {2023})}\BibitemShut {NoStop}%
\bibitem [{\citenamefont {Patel}\ \emph {et~al.}(2008)\citenamefont {Patel},
  \citenamefont {Markov},\ and\ \citenamefont
  {Hayes}}]{Patel_Markov_Hayes_2008}%
  \BibitemOpen
  \bibfield  {author} {\bibinfo {author} {\bibfnamefont {K.}~\bibnamefont
  {Patel}}, \bibinfo {author} {\bibfnamefont {I.}~\bibnamefont {Markov}},\ and\
  \bibinfo {author} {\bibfnamefont {J.}~\bibnamefont {Hayes}},\ }\bibfield
  {title} {\bibinfo {title} {Optimal synthesis of linear reversible circuits},\
  }\href {https://doi.org/10.26421/QIC8.3-4-4} {\bibfield  {journal} {\bibinfo
  {journal} {Quantum Information and Computation}\ }\textbf {\bibinfo {volume}
  {8}},\ \bibinfo {pages} {282–294} (\bibinfo {year} {2008})}\BibitemShut
  {NoStop}%
\bibitem [{\citenamefont {Schaeffer}\ and\ \citenamefont
  {Perkowski}(2014)}]{schaeffer2014costminimizationapproachsynthesis}%
  \BibitemOpen
  \bibfield  {author} {\bibinfo {author} {\bibfnamefont {B.}~\bibnamefont
  {Schaeffer}}\ and\ \bibinfo {author} {\bibfnamefont {M.}~\bibnamefont
  {Perkowski}},\ }\href {https://doi.org/10.48550/arXiv.1407.0070} {\bibinfo
  {title} {A cost minimization approach to synthesis of linear reversible
  circuits}} (\bibinfo {year} {2014}),\ \Eprint
  {https://arxiv.org/abs/1407.0070} {arXiv:1407.0070 [cs.ET]} \BibitemShut
  {NoStop}%
\bibitem [{\citenamefont {Webster}\ \emph {et~al.}(2025)\citenamefont
  {Webster}, \citenamefont {Koutsioumpas},\ and\ \citenamefont
  {Browne}}]{webster2025heuristicoptimalsynthesiscnot}%
  \BibitemOpen
  \bibfield  {author} {\bibinfo {author} {\bibfnamefont {M.}~\bibnamefont
  {Webster}}, \bibinfo {author} {\bibfnamefont {S.}~\bibnamefont
  {Koutsioumpas}},\ and\ \bibinfo {author} {\bibfnamefont {D.~E.}\ \bibnamefont
  {Browne}},\ }\href {https://arxiv.org/abs/2503.14660} {\bibinfo {title}
  {Heuristic and optimal synthesis of cnot and clifford circuits}} (\bibinfo
  {year} {2025}),\ \Eprint {https://arxiv.org/abs/2503.14660} {arXiv:2503.14660
  [quant-ph]} \BibitemShut {NoStop}%
\bibitem [{\citenamefont {Webster}\ \emph {et~al.}(2023)\citenamefont
  {Webster}, \citenamefont {Quintavalle},\ and\ \citenamefont
  {Bartlett}}]{Webster2023diagonallogical}%
  \BibitemOpen
  \bibfield  {author} {\bibinfo {author} {\bibfnamefont {M.~A.}\ \bibnamefont
  {Webster}}, \bibinfo {author} {\bibfnamefont {A.~O.}\ \bibnamefont
  {Quintavalle}},\ and\ \bibinfo {author} {\bibfnamefont {S.~D.}\ \bibnamefont
  {Bartlett}},\ }\bibfield  {title} {\bibinfo {title} {Transversal diagonal
  logical operators for stabiliser codes},\ }\href
  {https://doi.org/10.1088/1367-2630/acfc5f} {\bibfield  {journal} {\bibinfo
  {journal} {New Journal of Physics}\ }\textbf {\bibinfo {volume} {25}},\
  \bibinfo {pages} {103018} (\bibinfo {year} {2023})}\BibitemShut {NoStop}%
\bibitem [{\citenamefont {Bravyi}\ and\ \citenamefont
  {Kitaev}(2005)}]{bravyiUniversalQuantumComputation2005}%
  \BibitemOpen
  \bibfield  {author} {\bibinfo {author} {\bibfnamefont {S.}~\bibnamefont
  {Bravyi}}\ and\ \bibinfo {author} {\bibfnamefont {A.}~\bibnamefont
  {Kitaev}},\ }\bibfield  {title} {\bibinfo {title} {Universal quantum
  computation with ideal {{Clifford}} gates and noisy ancillas},\ }\href
  {https://doi.org/10.1103/PhysRevA.71.022316} {\bibfield  {journal} {\bibinfo
  {journal} {Phys. Rev. A}\ }\textbf {\bibinfo {volume} {71}},\ \bibinfo
  {pages} {022316} (\bibinfo {year} {2005})}\BibitemShut {NoStop}%
\bibitem [{\citenamefont {Fowler}\ and\ \citenamefont
  {Gidney}(2019)}]{fowlerLowOverheadQuantum2019}%
  \BibitemOpen
  \bibfield  {author} {\bibinfo {author} {\bibfnamefont {A.~G.}\ \bibnamefont
  {Fowler}}\ and\ \bibinfo {author} {\bibfnamefont {C.}~\bibnamefont
  {Gidney}},\ }\bibfield  {title} {\bibinfo {title} {Low overhead quantum
  computation using lattice surgery},\ }\href {http://arxiv.org/abs/1808.06709}
  {\bibfield  {journal} {\bibinfo  {journal} {arXiv}\ } (\bibinfo {year}
  {2019})}\BibitemShut {NoStop}%
\bibitem [{\citenamefont {Terhal}(2015)}]{terhalQuantumErrorCorrection2015}%
  \BibitemOpen
  \bibfield  {author} {\bibinfo {author} {\bibfnamefont {B.~M.}\ \bibnamefont
  {Terhal}},\ }\bibfield  {title} {\bibinfo {title} {Quantum error correction
  for quantum memories},\ }\href {https://doi.org/10.1103/RevModPhys.87.307}
  {\bibfield  {journal} {\bibinfo  {journal} {Reviews of Modern Physics}\
  }\textbf {\bibinfo {volume} {87}},\ \bibinfo {pages} {307} (\bibinfo {year}
  {2015})}\BibitemShut {NoStop}%
\bibitem [{\citenamefont {Fazio}\ \emph {et~al.}(2024)\citenamefont {Fazio},
  \citenamefont {Harper},\ and\ \citenamefont
  {Bartlett}}]{fazio2024logicalnoisebiasmagic}%
  \BibitemOpen
  \bibfield  {author} {\bibinfo {author} {\bibfnamefont {N.}~\bibnamefont
  {Fazio}}, \bibinfo {author} {\bibfnamefont {R.}~\bibnamefont {Harper}},\ and\
  \bibinfo {author} {\bibfnamefont {S.}~\bibnamefont {Bartlett}},\ }\href
  {https://arxiv.org/abs/2401.10982} {\bibinfo {title} {Logical noise bias in
  magic state injection}} (\bibinfo {year} {2024}),\ \Eprint
  {https://arxiv.org/abs/2401.10982} {arXiv:2401.10982 [quant-ph]} \BibitemShut
  {NoStop}%
\bibitem [{\citenamefont {Gokhale}\ \emph {et~al.}(2020)\citenamefont
  {Gokhale}, \citenamefont {Angiuli}, \citenamefont {Ding}, \citenamefont
  {Gui}, \citenamefont {Tomesh}, \citenamefont {Suchara}, \citenamefont
  {Martonosi},\ and\ \citenamefont {Chong}}]{gokhale2020optimization}%
  \BibitemOpen
  \bibfield  {author} {\bibinfo {author} {\bibfnamefont {P.}~\bibnamefont
  {Gokhale}}, \bibinfo {author} {\bibfnamefont {O.}~\bibnamefont {Angiuli}},
  \bibinfo {author} {\bibfnamefont {Y.}~\bibnamefont {Ding}}, \bibinfo {author}
  {\bibfnamefont {K.}~\bibnamefont {Gui}}, \bibinfo {author} {\bibfnamefont
  {T.}~\bibnamefont {Tomesh}}, \bibinfo {author} {\bibfnamefont
  {M.}~\bibnamefont {Suchara}}, \bibinfo {author} {\bibfnamefont
  {M.}~\bibnamefont {Martonosi}},\ and\ \bibinfo {author} {\bibfnamefont
  {F.~T.}\ \bibnamefont {Chong}},\ }\bibfield  {title} {\bibinfo {title}
  {Optimization of simultaneous measurement for variational quantum eigensolver
  applications},\ }in\ \href@noop {} {\emph {\bibinfo {booktitle} {2020 IEEE
  international conference on Quantum Computing and Engineering (QCE)}}}\
  (\bibinfo {organization} {IEEE},\ \bibinfo {year} {2020})\ pp.\ \bibinfo
  {pages} {379--390}\BibitemShut {NoStop}%
\bibitem [{\citenamefont {Cai}\ \emph {et~al.}(2023)\citenamefont {Cai},
  \citenamefont {Siegel},\ and\ \citenamefont
  {Benjamin}}]{caiLoopedPipelinesEnabling2023}%
  \BibitemOpen
  \bibfield  {author} {\bibinfo {author} {\bibfnamefont {Z.}~\bibnamefont
  {Cai}}, \bibinfo {author} {\bibfnamefont {A.}~\bibnamefont {Siegel}},\ and\
  \bibinfo {author} {\bibfnamefont {S.}~\bibnamefont {Benjamin}},\ }\bibfield
  {title} {\bibinfo {title} {Looped pipelines enabling effective {{3D}} qubit
  lattices in a strictly {{2D}} device},\ }\href
  {https://doi.org/10.1103/PRXQuantum.4.020345} {\bibfield  {journal} {\bibinfo
   {journal} {PRX Quantum}\ }\textbf {\bibinfo {volume} {4}},\ \bibinfo {pages}
  {20345} (\bibinfo {year} {2023})}\BibitemShut {NoStop}%
\bibitem [{\citenamefont {Gidney}(2025)}]{gidneyHowFactor20482025}%
  \BibitemOpen
  \bibfield  {author} {\bibinfo {author} {\bibfnamefont {C.}~\bibnamefont
  {Gidney}},\ }\bibfield  {title} {\bibinfo {title} {How to factor 2048 bit
  {{RSA}} integers with less than a million noisy qubits},\ }\bibfield
  {journal} {\bibinfo  {journal} {arXiv}\ }\href
  {https://doi.org/10.48550/arXiv.2505.15917} {10.48550/arXiv.2505.15917}
  (\bibinfo {year} {2025})\BibitemShut {NoStop}%
\bibitem [{\citenamefont {Brown}\ \emph {et~al.}(2017)\citenamefont {Brown},
  \citenamefont {Laubscher}, \citenamefont {Kesselring},\ and\ \citenamefont
  {Wootton}}]{brownPokingHolesCutting2017}%
  \BibitemOpen
  \bibfield  {author} {\bibinfo {author} {\bibfnamefont {B.~J.}\ \bibnamefont
  {Brown}}, \bibinfo {author} {\bibfnamefont {K.}~\bibnamefont {Laubscher}},
  \bibinfo {author} {\bibfnamefont {M.~S.}\ \bibnamefont {Kesselring}},\ and\
  \bibinfo {author} {\bibfnamefont {J.~R.}\ \bibnamefont {Wootton}},\
  }\bibfield  {title} {\bibinfo {title} {Poking holes and cutting corners to
  achieve clifford gates with the surface code},\ }\href
  {https://doi.org/10.1103/PhysRevX.7.021029} {\bibfield  {journal} {\bibinfo
  {journal} {Physical Review X}\ }\textbf {\bibinfo {volume} {7}},\ \bibinfo
  {pages} {21029} (\bibinfo {year} {2017})}\BibitemShut {NoStop}%
\end{thebibliography}
\end{document}